\newcommand{\al}{\mbox{$^{26}$\hspace{-0.2em}Al}}
\newcommand{\ti}{\mbox{$^{44}$Ti}}
\newcommand{\Msol}{\mbox{$M_{\sun}$}}
\newcommand{\Rsol}{\mbox{$R_{\sun}$}}

\newcommand{\pcmq}{\mbox{cm$^{-2}$}}

\newcommand{\psec}{\mbox{s$^{-1}$}}
\newcommand{\psr}{\mbox{sr$^{-1}$}}
\newcommand{\eunit}{\mbox{cm$^{2}$ s}}
\newcommand{\emunit}{\mbox{ph \psec}}
\newcommand{\peunit}{\mbox{\psec}}
\newcommand{\funit}{\mbox{ph \pcmq \psec}}
\newcommand{\fster}{\mbox{ph \pcmq \psec \psr}}

\newcommand{\mlr}{\mbox{MLR}}
\newcommand{\rmlr}{\mbox{RMLR}}
\newcommand{\dof}{\mbox{DOF}}
\newcommand{\um}{\mbox{$\mu $m}}
\newcommand{\lgamma}{\mbox{L$_{511}$}}
\newcommand{\lpositron}{\mbox{L$_{\rm p}$}}
\newcommand{\fpositron}{\mbox{$f_{\rm p}$}}
\newcommand{\dete}{\mbox{\tt DETE}}
\newcommand{\orbit}{\mbox{\tt ORBIT-DETE}}
\newcommand{\cface}{\mbox{$C_{\perp}$}}
\newcommand{\cedge}{\mbox{$C_{\parallel}$}}
\newcommand{\bgpars}{\mbox{\boldmath$\beta$}}
\def\MeV{\mbox{Me\hspace{-0.1em}V}}

\def\deg{\ensuremath{^\circ}}

\documentclass{aa}
\usepackage{epsfig}
\begin{document}

\title{The all-sky distribution of 511 keV electron-positron annihilation 
       emission\thanks{
	   Based on observations with INTEGRAL, an ESA project with instruments 
	   and science data centre funded by ESA member states (especially the 
	   PI countries: Denmark, France, Germany, Italy, Switzerland, Spain), 
	   Czech Republic and Poland, and with the participation of Russia and 
	   the USA.}}

\author{J.~Kn\"odlseder\inst{1}
         \and P.~Jean\inst{1}
         \and V.~Lonjou\inst{1}
         \and G.~Weidenspointner\inst{1}
	 \and N.~Guessoum\inst{2}
	 \and W.~Gillard\inst{1}
	 \and G.~Skinner\inst{1}
	 \and P.~von~Ballmoos\inst{1}
	 \and G.~Vedrenne\inst{1}
	 \and J.-P.~Roques\inst{1}
	 \and S.~Schanne\inst{3}
	 \and B.~Teegarden\inst{4}
	 \and V.~Sch\"onfelder\inst{5}
	 \and C.~Winkler\inst{6}
}

\offprints{J\"urgen Kn\"odlseder, e-mail : knodlseder@cesr.fr}

\institute{
$^{1}$ Centre d'\'Etude Spatiale des Rayonnements, CNRS/UPS, B.P.~4346, 
	   31028 Toulouse Cedex 4, France \\
$^{2}$ American University of Sharjah, College of Arts \& Science,
       Physics Department, PO Box 26666, Sharjah, UAE \\
$^{3}$ DSM/DAPNIA/SAp, CEA Saclay, 91191 Gif-sur-Yvette, France \\
$^{4}$ Laboratory for High Energy Astrophysics, NASA/Goddard Space Flight 
       Center, Greenbelt, MD 20771, USA \\
$^{5}$ Max-Planck-Institut f\"ur Extraterrestrische Physik, Postfach 1603, 
       85740 Garching, Germany \\
$^{6}$ ESA/ESTEC, Science Operations and Data Systems Division 
       (SCI-SD), 2201 AZ Noordwijk, The Netherlands
}

\date{Received / Accepted }

\authorrunning{J.~Kn\"odlseder et al.}

\titlerunning{The all-sky distribution of 511 keV electron-positron annihilation 
              emission}

\abstract{
We present a map of 511 keV electron-positron annihilation emission, 
based on data accumulated with the SPI spectrometer aboard ESA's INTEGRAL
gamma-ray observatory,
that covers approximately $\sim95\%$ of the celestial sphere.
Within the exposed sky area, 511 keV line emission is significantly 
detected towards the galactic bulge region and, at a very low level, 
from the galactic disk.
The bulge emission is highly symmetric and is centred on the galactic 
centre with an extension of $\sim 8\deg$ (FWHM).
The emission is equally well described by models that represent the 
stellar bulge or halo populations.
The detection significance of the bulge emission is $\sim 50\sigma$, 
that of the galactic disk is $\sim 4\sigma$.
The disk morphology is only weakly constrained by the present data, 
being compatible with both the distribution of young and old stellar 
populations.
The 511~keV line flux from the bulge and disk components is
$(1.05 \pm 0.06) \times 10^{-3}$ \funit\ and
$(0.7 \pm 0.4) \times 10^{-3}$ \funit, respectively, 
corresponding to a bulge-to-disk flux ratio in the range $1-3$.
Assuming a positronium fraction of $\fpositron=0.93$ this translates 
into annihilation rates of
$(1.5 \pm 0.1) \times 10^{43}$ \peunit\ and 
$(0.3 \pm 0.2) \times 10^{43}$ \peunit, respectively.
The ratio of the bulge luminosity to that of the disk is in the range $3-9$.
We find no evidence for a point-like source in addition to the 
diffuse emission, down to a typical flux limit of $\sim10^{-4}$ \funit.
We also find no evidence for the positive latitude enhancement that 
has been reported from OSSE measurements; our $3\sigma$ upper flux limit 
for this feature is $1.5 \times 10^{-4}$ \funit.
The disk emission can be attributed to the $\beta^+$-decay of the 
radioactive species \al\ and \ti.
The bulge emission arises from a different source which has only a 
weak or no disk component.
We suggest that Type Ia supernovae and/or low-mass X-ray binaries are the 
prime candidates for the source of the galactic bulge positrons.
Light dark matter annihilation could also explain the observed 
511~keV bulge emission characteristics.
\keywords{gamma rays: observations -- line: profiles -- Galaxy: centre}
}

\maketitle

\section{Introduction}

Since the first detection (Johnson \& Haymes \cite{johnson73}) and 
the subsequent firm identification (Leventhal et al. \cite{leventhal78})
of the galactic 511 keV annihilation line, the origin of galactic 
positrons has been a lively topic of scientific debate.
Among the proposed candidates for sources of positrons figure
cosmic-ray interactions with the interstellar medium 
(Ramaty et al.~\cite{ramaty70}),
pulsars (Sturrock \cite{sturrock71}),
compact objects housing either neutron stars or black holes
(Ramaty \& Lingenfelter \cite{ramaty79}),
gamma-ray bursts (Lingenfelter \& Hueter \cite{lingenfelter84}),
(light) dark matter 
(Rudaz \& Stecker \cite{rudaz88}; Boehm et al.~\cite{boehm04}),
and stars expelling radioactive nuclei produced by nucleosynthesis, 
such as 
supernovae (Clayton \cite{clayton73}), 
hypernovae (Cass\'e et al.~\cite{casse04}), 
novae (Clayton \& Hoyle \cite{clayton74}),
red giants (Norgaard \cite{norgaard80}),
and Wolf-Rayet stars (Dearborn \& Blake \cite{dearborn85}).
It seems difficult to disentangle the primary galactic positron source 
based only on theoretical grounds, mainly due to the (highly) uncertain 
positron yields, but also due to the uncertain distribution and duty 
cycle of the source populations.

Help is expected from a detailed study of the 511 keV line emission morphology.
The celestial 511 keV intensity distribution should be tied to the 
spatial source distribution, although positron diffusion and effects 
associated with the annihilation physics may to some extent blur this link.
First estimations of the 511 keV emission morphology were obtained 
by the Oriented Scintillation Spectrometer Experiment (OSSE) on-board 
the Compton Gamma-Ray Observatory (CGRO) satellite
(Purcell et al. \cite{purcell94}; 
Cheng et al. \cite{cheng97}; 
Purcell et al. \cite{purcell97};
Milne et al. \cite{milne00}; 
Milne et al. \cite{milne01}),
but observations were restricted to the inner Galaxy, giving only 
a limited view of the 511 keV emission distribution.
With the launch of ESA's INTEGRAL satellite in October 2002, a new 
gamma-ray observatory is available that allows a detailed study 
of positron annihilation signatures.
In particular, the imaging spectrometer SPI 
(Vedrenne et al.~\cite{vedrenne03}), 
one of the two prime instruments on-board INTEGRAL, has been optimised 
for the study of line radiation, combining 
high-resolution spectroscopy (R $\sim250$ at 511 keV) with modest 
angular resolution ($3\deg$ FWHM).

We present in this work an all-sky map of 511 keV gamma-ray line 
emission, with the goals of determining the morphology of 
the emission in the Galaxy and of searching for previously unknown 
sources of 511~keV emission anywhere in the sky.
The present public data archive does not yet cover the entire 
celestial sphere, but the unexposed regions are limited to a few 
areas at high galactic latitudes, comprising less than $5\%$ of the sky.The resulting point-source sensitivity is better than 
$2\times10^{-4}$ \funit\ for many regions along the galactic 
plane, allowing for the first time the extraction of information about the 
distribution of positron annihilation all over the Galaxy.
We do not address the distribution of positronium continuum emission 
in this paper, since the subtraction of the diffuse galactic continuum 
emission is a distinct data analysis challenge.
A map of positronium continuum emission will be presented elsewhere
(Weidenspointner et al., in preparation).

Earlier results on the 511~keV line emission morphology as observed by 
SPI have been presented by
Jean et al.~(\cite{jean03a}),
Kn\"odlseder et al.~(\cite{knoedl04a}), and
Weidenspointner et al.~(\cite{weidenspointner04}),
and were based on observations performed during the galactic centre 
deep exposure (GCDE) of 2003.
Using a ``light bucket'' approach which neglects the coding 
properties of the SPI mask,
Teegarden et al.~(\cite{teegarden05}) derived upper limits on 
electron-positron annihilation radiation from the galactic disk using 
core-programme data combined with open-programme observations at low 
galactic latitudes ($|b| \le 20\deg$).
In the present paper we provide for the first time an all-sky analysis 
using all public data of the first INTEGRAL mission year.

Spectroscopic characteristics of the 511~keV line based on SPI data have 
been published by
Jean et al.~(\cite{jean03a}),
Lonjou et al.~(\cite{lonjou04}), and
Churazov et al.~(\cite{churazov05}).
We will present the 511~keV line profile that we obtain from the 
all-sky dataset elsewhere (Jean et al., in preparation). 

This paper is organised as follows.
Section \ref{sec:obs} describes the observations and the 
data preparation.
Section \ref{sec:bgm} explains the treatment of the instrumental 
background.
In section \ref{sec:results}, we present the first all-sky map of 
511 keV gamma-ray line radiation and determine the morphology
of the emission.
Section \ref{sec:results} also describe searches for correlations with 
tracers of galactic source populations in order to shed light on the 
origin of the positrons.
In section \ref{sec:discussion} we discuss the implications of the 
observations for the galactic origin of positrons, and we conclude in 
section \ref{sec:conclusions}.

\section{Observations and Data Preparation}
\label{sec:obs}

The data that were analysed in this work consist of those included 
in the December 10, 2004 public INTEGRAL data release
(i.e.~orbital revolutions 19-76, 79-80, 89-122)
plus the INTEGRAL Science Working Team data of the Vela region observed 
during revolutions 81-88.
The data span the IJD epoch $1073.394-1383.573$, where IJD is the 
Julian Date minus 2\,451\,544.5 days.

We screened the data for anomalously high counting rates (typically 
occurring at the beginning and the end of an orbital revolution due 
to the exit and entry of the radiation belts) and 
for periods of solar activity (as monitored by the SPI anticoincidence 
system) and excluded these periods from the data.
This data screening has turned out to be crucial for reducing the systematic 
uncertainties in the data analysis related to instrumental background 
variations.
After data screening, the dataset consists of 6821 pointed 
observations, with a total exposure time of 15.3~Ms.
Typical exposure times per pointing are $1200-3400$ seconds, but 
a few long staring observations of up to 113 ks exposure time 
are also included.

\begin{figure*}[!t]
  \center
  \epsfxsize=18cm \epsfclipon
  \epsfbox{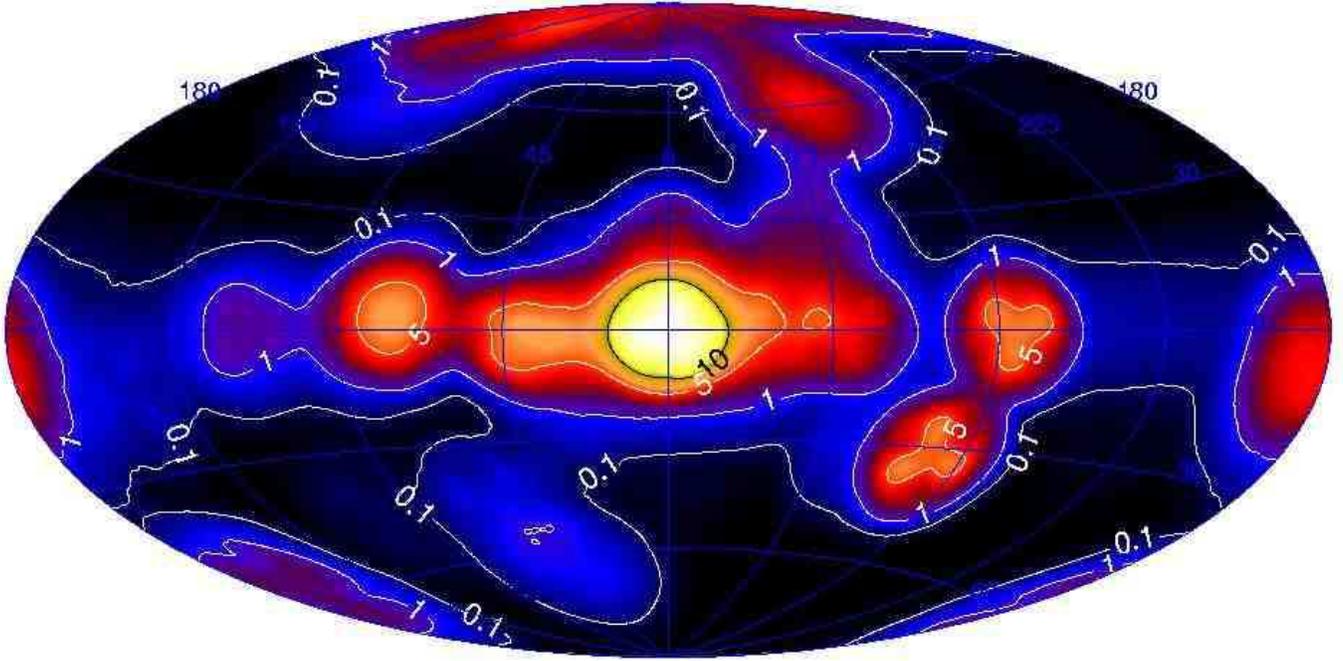}
  \caption{\label{fig:exposure}
  Map of the effective SPI exposure at 511 keV for the dataset 
  analysed in this work.
  The contours are labelled in units of $10^7$ \eunit, corresponding to
  13 ks (0.1), 133 ks (1), 667 ks (5), and 1.3 Ms (10) of effective exposure 
  times.
  }
\end{figure*}

Figure \ref{fig:exposure} shows a map of the resulting effective SPI exposure 
at 511 keV.
The maximum exposure of $2.1 \times 10^8$ \eunit\ occurs towards the 
galactic centre region thanks to data obtained during a long dedicated 
observation of this region.\footnote{
  To obtain the effective exposure time, the exposure has to be divided
  by the effective area at 511 keV of about 75 cm$^2$.
  }
A relatively uniform exposure of $\sim 3\times10^7$ \eunit\ has been 
achieved for galactic longitudes $|l| \le 50\deg$ and latitudes
$|b| \le 15\deg$.
Regions of peculiarly high exposure ($\sim 5\times10^7$ \eunit) 
are found in Cygnus, Vela and towards the Large Magellanic Cloud.
In addition, particularly well exposed sources ($\ga 2\times10^7$ \eunit) 
are the Crab nebula, 3C~273, NGC~4151, M~94, NGC~936 
(during the SN2003~gs outburst) and the Coma cluster.
Unexposed regions are found mostly at intermediate galactic latitudes
($|b| \sim 30\deg-60\deg$), and towards the south galactic pole.

\begin{figure}[!t]
  \center
  \epsfxsize=8.8cm \epsfclipon
  \epsfbox{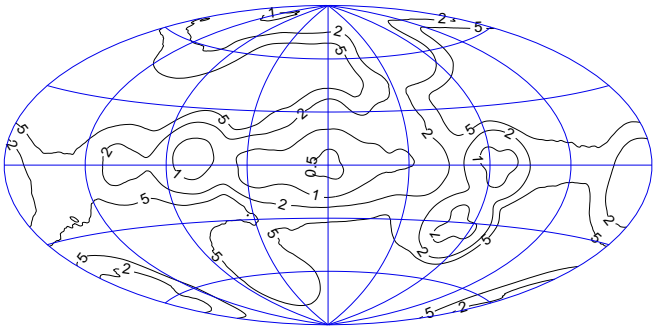}
  \caption{\label{fig:sensitivity}
  SPI narrow line $3\sigma$ point-source sensitivity at 511~keV for the 
  analysis interval $507.5-514.5$ keV 
  (contours are labelled in units of $10^{-4}$ \funit).
  }
\end{figure}

A map of the resulting narrow-line $3\sigma$ point-source 
sensitivity of SPI at 511 keV is shown in Fig.~\ref{fig:sensitivity}.
To evaluate the sensitivity, an energy band of 7~keV centred 
at 511~keV has been used.
The choice of such a relatively wide band eliminates any bias due to 
the germanium detector degradation and annealing cycles, as well as 
any bias/effect due to gain calibration uncertainties.
It also takes into account moderate 511~keV line broadening, as 
reported by Jean et al.~(\cite{jean03a}).

Over large regions of the sky, and in particular in the galactic 
plane, a sensitivity better than $2\times10^{-4}$ \funit\ is reached.
A best point-source sensitivity of $5\times10^{-5}$ \funit\ 
is achieved towards the galactic centre direction.
The sensitivity to extended diffuse emission becomes slightly worse 
with increasing emission size, and depends on the exposure pattern in 
the region of interest.
For example, for a 2d angular Gaussian surface brightness distribution
centred on the galactic centre, the 511~keV line sensitivity worsens from
$5\times10^{-5}$ \funit\ for a galactic centre point-source to
$7\times10^{-5}$ \funit\ for an extended source of $8\deg$ FWHM.

Only single-detector event data have been analysed in this work 
(multiple-detector event data do not contribute significantly to the 
SPI sensitivity at an energy of 511~keV; c.f.~Roques et al.~\cite{roques03}).
Energy calibration was performed orbit-wise, resulting in a relative 
(orbit-to-orbit) calibration precision of $\sim0.01$ keV and an absolute 
accuracy of $\sim0.05$ keV (Lonjou et al.~\cite{lonjou04}).

The data have been analysed by sorting the events in a 3-dimensional 
data-space, spanned by the (calibrated) event energy, the detector 
number, and the SPI pointing number.
An energy binning of 0.5 keV has been chosen, well below the 
instrumental energy resolution of $2.12$~keV at 511 keV.

\section{Background modelling}
\label{sec:bgm}

The most crucial step in SPI data analysis consists of the precise 
modelling of the time variability of the instrumental background.
In the region of the 511~keV line, the instrumental background consists of 
a nearly flat continuum and a (broadened) instrumental 511~keV line 
originating from positron annihilation within the telescope 
(Teegarden et al.~\cite{teegarden04}).
Since the time variation of the continuum component differs from that of the
line component we model them independently.
The background model for a given data-space bin, indexed in the following by 
the pointing number $p$, the detector number $d$ and the energy bin $e$, 
is then given by
\begin{equation}
 b_{p,d,e} = b^{\rm cont}_{p,d,e} + b^{\rm line}_{p,d,e}
 \label{eq:bgm}
\end{equation}
(note that for the analysis presented in this work a single energy 
bin has been used, covering the energy interval $507.5-514.5$ keV;
however, for clarity and reference in future works we give here the 
complete energy-dependent formalism).

The time variation of the continuum component is extrapolated from 
that observed in an continuum energy band adjacent to the 511~keV line.
We used the energy band $E_{\rm adj} = 523-545$~keV, situated above the 
511~keV line, in order to exclude any bias due to positronium continuum 
emission that appears below 511~keV.
To reduce the statistical uncertainty that arises from the limited 
counting statistics, we smoothed the time variation by 
locally adjusting the rate of saturated events in the germanium detectors 
(GEDSAT) to the adjacent counting rate
(GEDSAT turned out to provide a good first order tracer of the background 
variation in SPI; c.f.~Jean et al.~\cite{jean03b}). 
The predicted number of continuum background counts in data-space bin
$(p,d,e)$ is then given by
\begin{eqnarray}
 b^{\rm cont}_{p,d,e} & = & g_{p,d} \times T_{p,d} \times \nonumber \\
 & & \frac{\Delta_{e}}{\sum_{e' \in E_{\rm adj}} \Delta_{e'}}
     \times
     \frac{\sum_{p'=p-\Delta_{p}}^{p+\Delta_{p}} \sum_{e' \in E_{\rm adj}} 
	   n_{p',d,e'}}
          {\sum_{p'=p-\Delta_{p}}^{p+\Delta_{p}}  
           g_{p',d} \times T_{p',d}} ,
 \label{eq:contbgm}
\end{eqnarray}
where
\begin{itemize}
\item $g_{p,d}$ is the GEDSAT rate for detector $d$, averaged over the time 
      period spanned by pointing $p$, given in units of counts 
      s$^{-1}$,
\item $T_{p',d}$ is the lifetime for detector $d$ during pointing 
      $p'$, given in units of seconds,
\item $\Delta_{e}$ is the energy bin size for spectral bin $e$, given 
      in units of keV (here $\Delta_{e} = 0.5$ keV), and
\item $n_{p',d,e'}$ is the number of observed counts for pointing $p'$, 
      detector $d$, and energy bin $e'$, given in units of counts.
\end{itemize}
The number of pointings used for smoothing, given by 
$2 \Delta_{p} + 1$, is 
determined for each pointing $p$ and detector $d$ by satisfying the constraint
\begin{equation}
 \min_{\Delta_{p} \ge 0} \left(
   \sum_{p'=p-\Delta_{p}}^{p+\Delta_{p}} T_{p',d} \ge T_{\rm min}
   \right) .
\end{equation}
An accumulated lifetime of $T_{\rm min} = 20$ hours has shown to provide an 
optimum compromise between reducing the statistical uncertainty (due to the 
limited number of events in the adjacent energy band) and reducing the 
systematic uncertainty (due to the fact that the GEDSAT rate does not predict 
the background to infinite precision).
In other words, continuum background variations shorter than $\sim20$ hours 
are modelled by the GEDSAT rate while variations on longer time scales are 
modelled by the observed event rate in the $523-545$~keV band.

The time variation of the line component was modelled for each 
detector $d$ and energy bin $e$ separately using a multi-component template 
of the form
\begin{equation}
b^{\rm line}_{p,d,e} =
 \beta^{(1)}_{d,e} + 
 \beta^{(2)}_{d,e} \times g_{p,d} +
 \beta^{(3)}_{d,e} \int_{t_{0}}^t g_{d}(t') e^{(t'-t)/\tau} {\rm d}t' .
 \label{eq:linebgm}
\end{equation}

This template consists of a constant term $\beta^{(1)}_{d,e}$ plus
the GEDSAT rate $g_{p,d}$ scaled by $\beta^{(2)}_{d,e}$ 
plus the GEDSAT rate $g_{d}(t')$ convolved with an exponential 
decay law, scaled by $\beta^{(3)}_{d,e}$ (the convolution integral is taken 
from the start of the INTEGRAL mission $t_{0}$ up to date $t$).
The coefficients $\beta^{(i)}_{d,e}$ of the template are adjusted during the 
analysis for each SPI detector $d$ and energy bin $e$ using a maximum 
likelihood fitting procedure (again, in the analysis presented in 
this paper only a single energy bin is used).
The constant term has been introduced to provide for non-linearities 
between the background variation and the GEDSAT rate.
In fact, it turns out that $\beta^{(1)}_{d,e}$ are negative.
An equally good background predictor is obtained if the GEDSAT rate 
raised to a power of $\sim1.1$ is taken, but using a constant instead of a
powerlaw has the advantage of having the background variation template 
decomposed into a linear combination of terms.
The third component makes provision for a long term build-up that is 
seen in the intensity of the 511 keV background line, and that is 
tentatively attributed to production of the isotope $^{65}$Zn which 
has a decay time of $\tau=352$ days.
The precise value of $\tau$ is in fact weakly constrained by the 
present data, and a linear slope provides an equally good fit of the 
instrumental 511~keV line background.

\begin{figure}[!t]
  \center
  \epsfxsize=8.8cm \epsfclipon
  \epsfbox{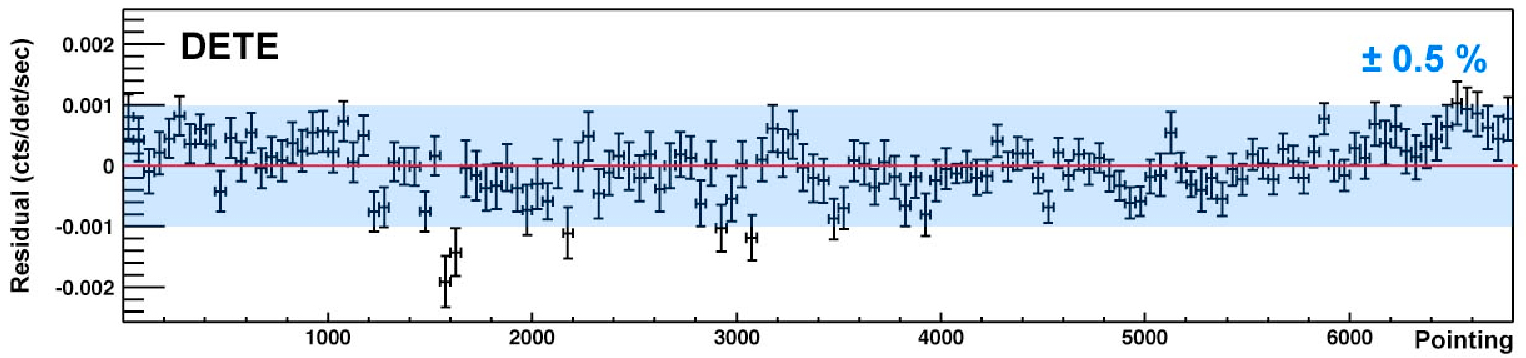}
  \epsfxsize=8.8cm \epsfclipon
  \epsfbox{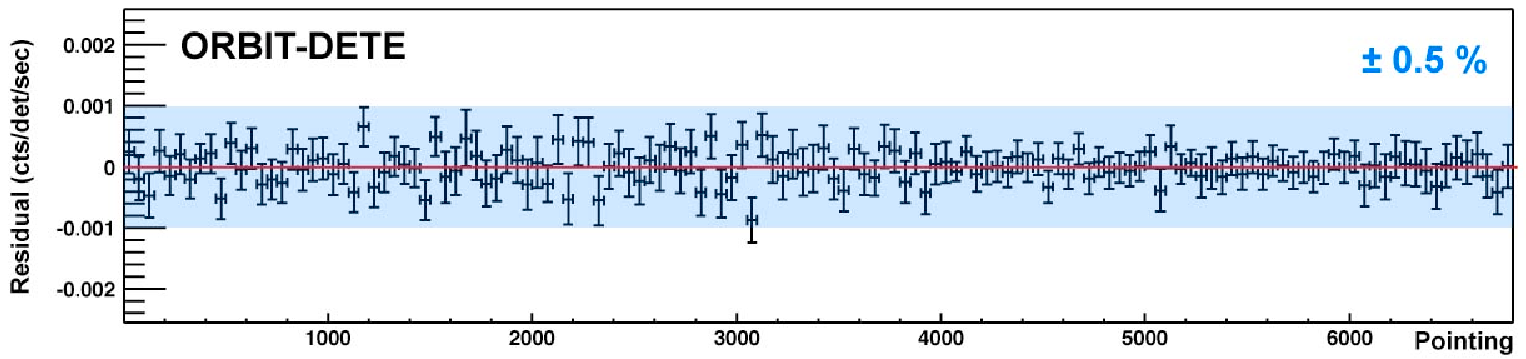}
  \caption{\label{fig:residual}
  Residual count rate as function of pointing number for the \dete\ 
  (top) and \orbit\ (bottom) background models for the energy band
  $507.5-514.5$ keV.
  In addition to the background models the best fitting 2d Gaussian 
  surface brightness model (c.f.~section \ref{sec:gaussian})
  has been subtracted from the data.
  For clarity, the data have been rebinned into groups of 50 pointings.
  The shaded area indicates count rate variations of $\pm0.5\%$.
  For comparison, the maximum 511~keV line signal amplitude corresponds 
  to approximately $\sim2\%$ of the observed count rate.
  }
\end{figure}

Although the background model defined by 
Eqs.~(\ref{eq:bgm})--(\ref{eq:linebgm}), 
which hereafter is called model \dete, predicts the 
instrumental background to good accuracy, significant residuals 
remain after subtracting off the background model and a model of the sky 
intensity distribution from the data (c.f.~Fig.~\ref{fig:residual}).
We found that these residuals can lead to systematic biases in the 
study of the morphology of the 511~keV emission, in 
particular for the determination of the longitude profile of the 
emission.
These biases can be explained by the telescope pointing strategy that has 
been adopted for a large fraction of the galactic centre deep exposure (GCDE) 
of the INTEGRAL core program: 
a slow ($5\deg$/day) scan of the galactic plane from 
negative towards positive longitudes combined with rapid ($3\deg$/hour)
excursions in galactic latitude.
As result, the longitude profile of the 511~keV line emission is 
encoded in count rate variations on timescales of days while the 
latitude profile is encoded in count rate variations on timescales 
of hours.
Within a few hours the SPI instrumental background is sufficiently 
stable to be accurately predicted by our model, hence the latitude 
profile is rather well determined.
However, on timescales of days the background variations are more
difficult to predict to sufficient accuracy, potentially leading to 
systematic trends in the determination of the longitude profile.\footnote{
  In a preliminary analysis in which we treated a much smaller 
  dataset, systematic background uncertainties suggested a significant 
  elongation of the galactic centre bulge emission along the galactic 
  plane. This elongation was artificial and had been 
  produced by background variations that were not fully explained by 
  our model. Removing the short period of data with the strongest 
  background variations removed also the apparent elongation of the bulge 
  emission.}

In order to improve the background model on long timescales, we studied 
also a class of models where we adjust the longterm variations during 
model fitting.
For this purpose we adjust the model parameters $\beta^{(2)}_{d,e}$ not only 
for all SPI detectors but also for time intervals of fixed duration $T$.
In this way, systematic uncertainties in the background model on 
timescales longer than $T$ are removed.
Fitting the background for each orbital revolution ($T\sim3$ days)
is adequate to reduce systematic trends well below the statistical 
uncertainties (c.f.~bottom panel of Fig.~\ref{fig:residual}).
This method is similar to the method that we applied in our earlier 
works 
(Jean et al.~\cite{jean03a};
Kn\"odlseder et al.~\cite{knoedl04a}; 
Weidenspointner et al.~\cite{weidenspointner04}), 
with the difference that we now also fit the background model for each of the 
SPI detectors separately, and that we included in addition a 
constant term and a build-up term in the model (see Eq.~\ref{eq:linebgm}).
Hereafter this second background model is called \orbit.

The introduction of additional parameters in \orbit\ with respect to 
\dete\ leads to a substantial loss in sensitivity.
The detection significance of galactic centre 511~keV line emission 
drops from $\sim50\sigma$ for \dete\ to $\sim22\sigma$ for \orbit.
However, it was found that the statistical accuracy of the morphology 
determination, which is driven by the count rate contrast in the 
data-space rather than the count rate level, is not degraded by the 
introduction of additional parameters, as long as $T \ga 2$ days.
Consequently, using the \orbit\ model for the morphological 
characterisation of the 511~keV line emission is the optimum choice 
that keeps a high statistical accuracy while reducing the systematic 
uncertainties in the analysis.

On the other hand, despite the systematic uncertainties,
\dete\ is accurate enough to allow for a precise determination of 511~keV 
line flux levels.
This is related to the fact that flux measurements require an average 
determination of the count rate level and are not sensitive to the 
count rate contrast.
Apparently, the count rate residuals approximately average to zero 
(c.f.~Fig.~\ref{fig:residual}).

We therefore opted for a two step approach where we first determine 
the morphology using \orbit, and then, using the optimum morphology 
parameters, determine the 511~keV flux using \dete.
In this way we recover the good sensitivity of SPI for 511~keV flux 
measurements that was reduced by a factor of $\sim2$ by the usage of 
\orbit.
The comparison of the flux levels determined using \dete\ and \orbit\ 
provides us with a measure of the systematic uncertainty in the flux 
determination, which in general is smaller than the statistical 
uncertainty obtained with \dete.
We add the systematic to the statistical uncertainty in quadrature 
and quote the result as total error on the flux measurement.
In cases where uncertainties in the morphology (such as the size of 
the emission region) introduce some uncertainty on the flux, we have 
also added this uncertainty to the total error in quadrature.

\section{Results}
\label{sec:results}

\subsection{Imaging}
\label{sec:map}

To determine a model independent map of the 511 keV gamma-ray line 
intensity distribution over the sky, we employed the Richardson-Lucy 
algorithm (Richardson \cite{richardson72}; Lucy \cite{lucy74}).
This type of algorithm is widely used for image deconvolution, and 
has in particular been successfully employed for the analysis of 
gamma-ray data of CGRO 
(Kn\"odlseder et al.~\cite{knoedl99a}; Milne et al.~\cite{milne00}).

\begin{figure*}[!t]
  \center
  \epsfxsize=18cm \epsfclipon
  \epsfbox{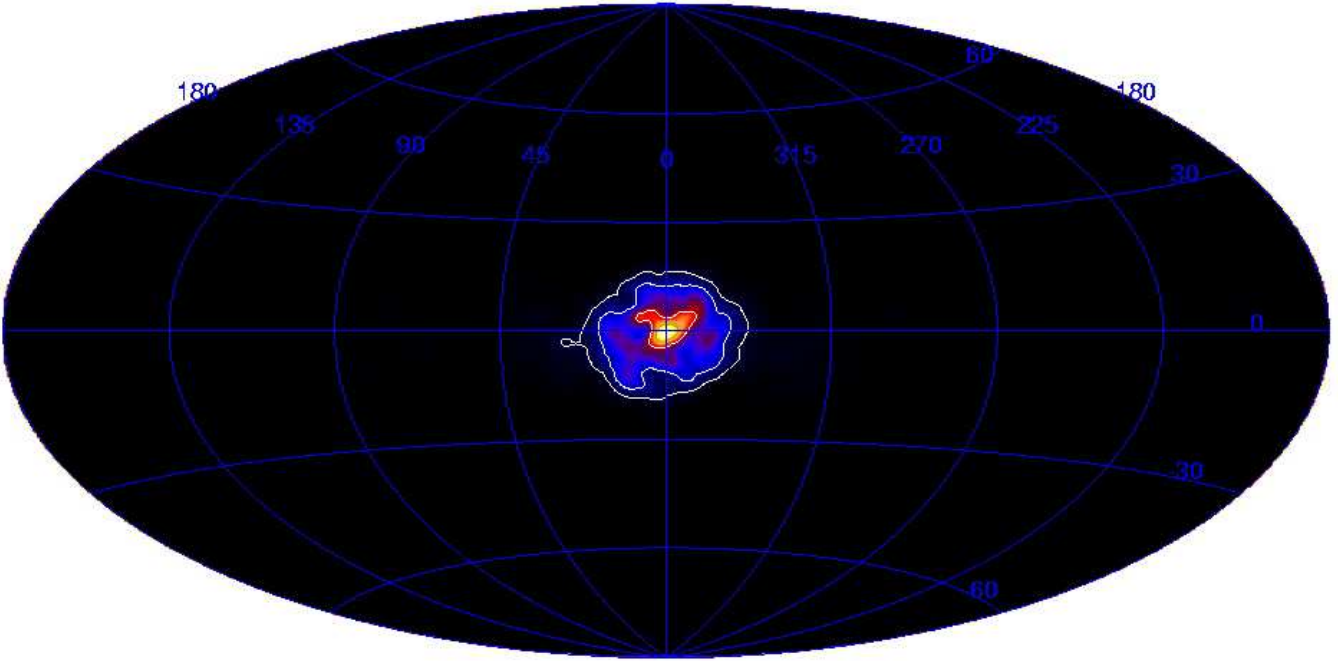}
  \caption{\label{fig:em}
  Richardson-Lucy image of 511~keV gamma-ray line emission (iteration 17).
  Contour levels indicate intensity levels of $10^{-2}$, $10^{-3}$, and 
  $10^{-4}$ \fster\ (from the centre outwards).
  }
\end{figure*}

\begin{figure*}[!t]
  \center
  \epsfxsize=8.8cm \epsfclipon
  \epsfbox{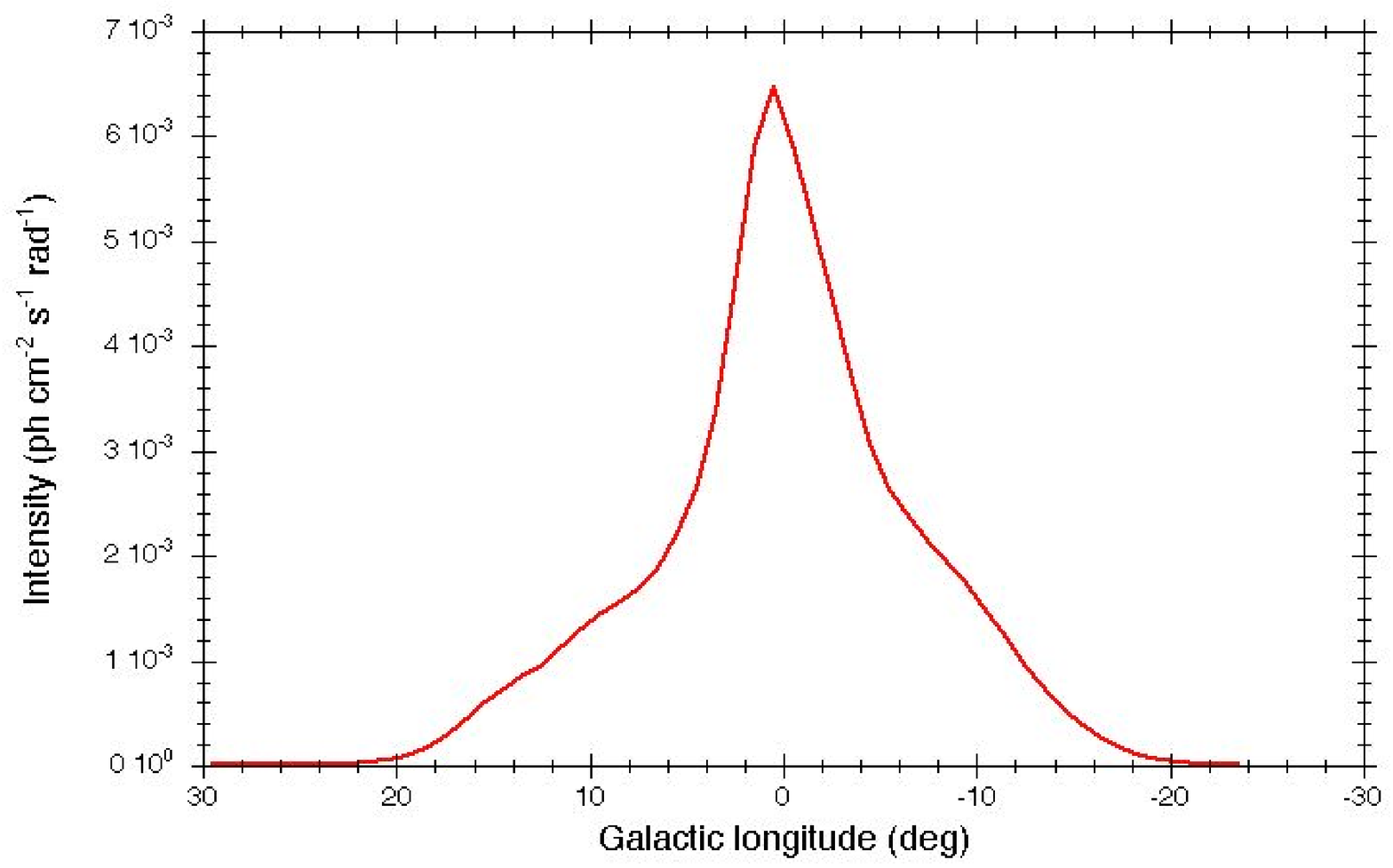}
  \hfill
  \epsfxsize=8.8cm \epsfclipon
  \epsfbox{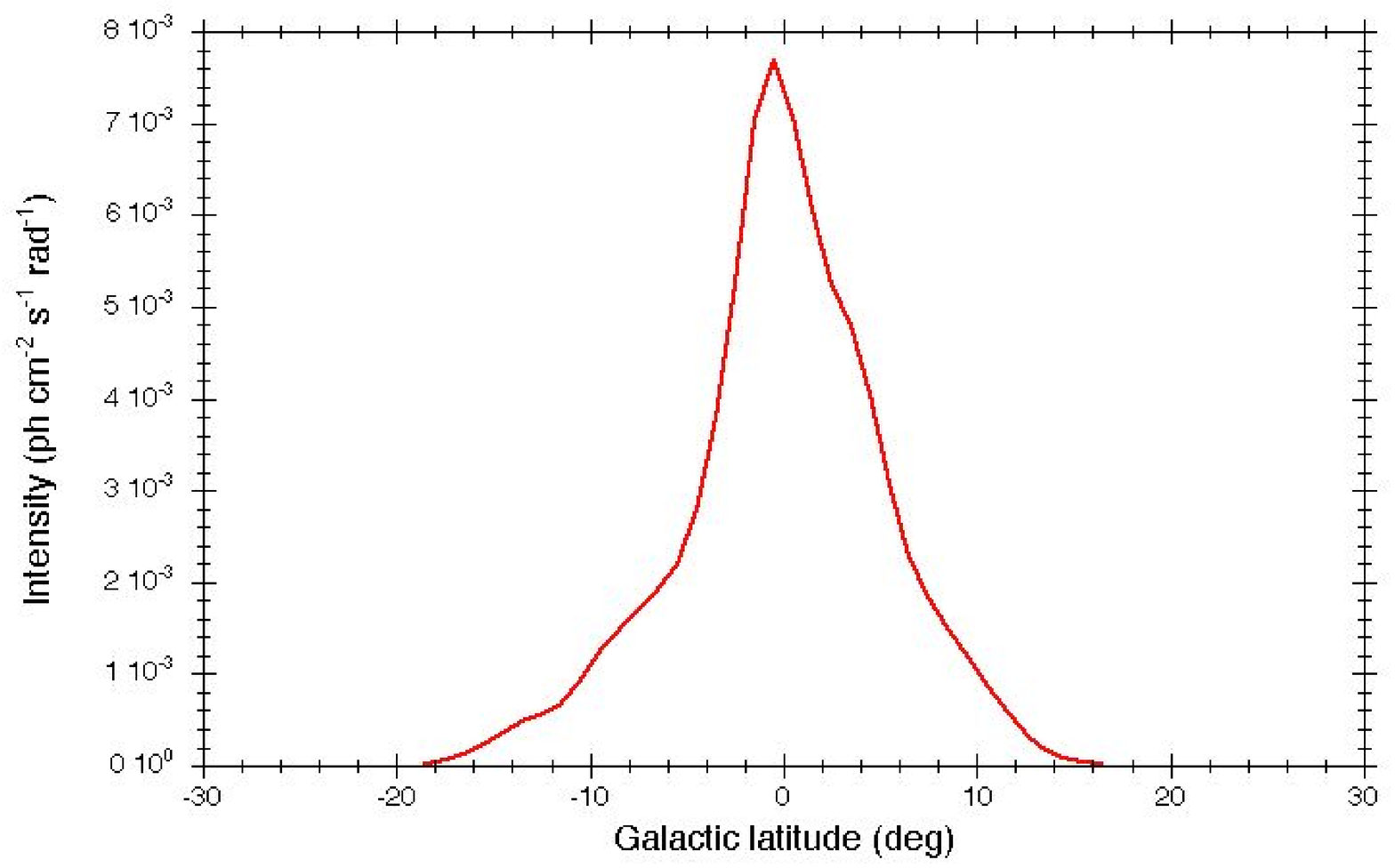}
  \caption{\label{fig:emprofiles}
  Longitude and latitude profiles of the image shown in 
  Fig.~\ref{fig:em} (integration range $|l| \le 30\deg$, $|b| \le 30\deg$).
  }
\end{figure*}

We implemented the accelerated version ML-LINB-1 of  
Kaufman (\cite{kaufman87}) of the Richardson-Lucy algorithm for our analysis, 
which iteratively updates the sky intensity distribution 
$f_j^k \to f_j^{k+1}$ using the relation
\begin{equation}
 f_j^{k+1} = f_j^k + \lambda^k w_j f_j^k 
             \left(\frac{\sum_{i=1}^N \left( \frac{n_{i}}{e_{i}^k} - 1 
                                                               \right) R_{ij}}
                        {\sum_{i=1}^N R_{ij}} \right)
\label{eq:em}
\end{equation}
where
$R_{ij}$ is the instrumental response matrix (linking the data space, 
indexed by $i$, to the image space, indexed by $j$),
$n_{i}$ is the number of counts measured in data space bin $i$,
$e_{i}^k = \sum_{j=1}^M R_{ij} f_j^k + b_{i}$ is the predicted number 
of counts in data space bin $i$ after iteration $k$ ($b_{i}$ 
being the predicted number of instrumental background counts for bin 
$i$),
$N$ and $M$ are the dimensions of the data and image space, 
respectively, and $\lambda^k$ is an acceleration factor that is 
obtained by constrained maximum likelihood fitting (with the 
constraint that the resulting sky intensities remain positive).

To avoid noise artefacts in the weakly exposed regions of the sky, we 
weighted the image increment with a quantity that is related to
the sensitivity of the instrument, given by 
$w_j = (\sum_{i=1}^N R_{ij})^{1/2}$.
We verified that introducing this weighting had no impact on the image 
reconstruction in the well exposed regions of the sky.
In addition, we smoothed the iterative corrections on the right hand side of 
Eq.~\ref{eq:em} using a $5\deg \times 5\deg$ boxcar average.
In this way the effective number of free parameters in the 
reconstruction is reduced and image noise is damped to an acceptable 
level.
The application of more sophisticated image reconstruction methods 
involving wavelet based multi-resolution algorithms aiming at a 
complete suppression of image noise (Kn\"odlseder et al.~\cite{knoedl99a})
will be presented elsewhere.

The resulting all-sky image of the 511~keV line emission is shown in 
Fig.~\ref{fig:em}, longitude and latitude profiles of the emission 
are shown in Fig.~\ref{fig:emprofiles}.
We have chosen to stop the iterative procedure after iteration 17 
since at this point the recovered flux and the fit quality correspond 
approximately to the values that we achieve by fitting 
astrophysical models to the data (c.f.~section \ref{sec:morphology}).
In this way we make sure that we are not in the regime of overfitting, 
which is characterised by substantial image noise and artificial image 
structures.
On the other hand, simulations showed that faint diffuse emission, as 
expected for example for a galactic disk component, would not be 
recovered at this point.

\begin{figure*}[!t]
  \center
  \epsfxsize=18cm \epsfclipon
  \epsfbox{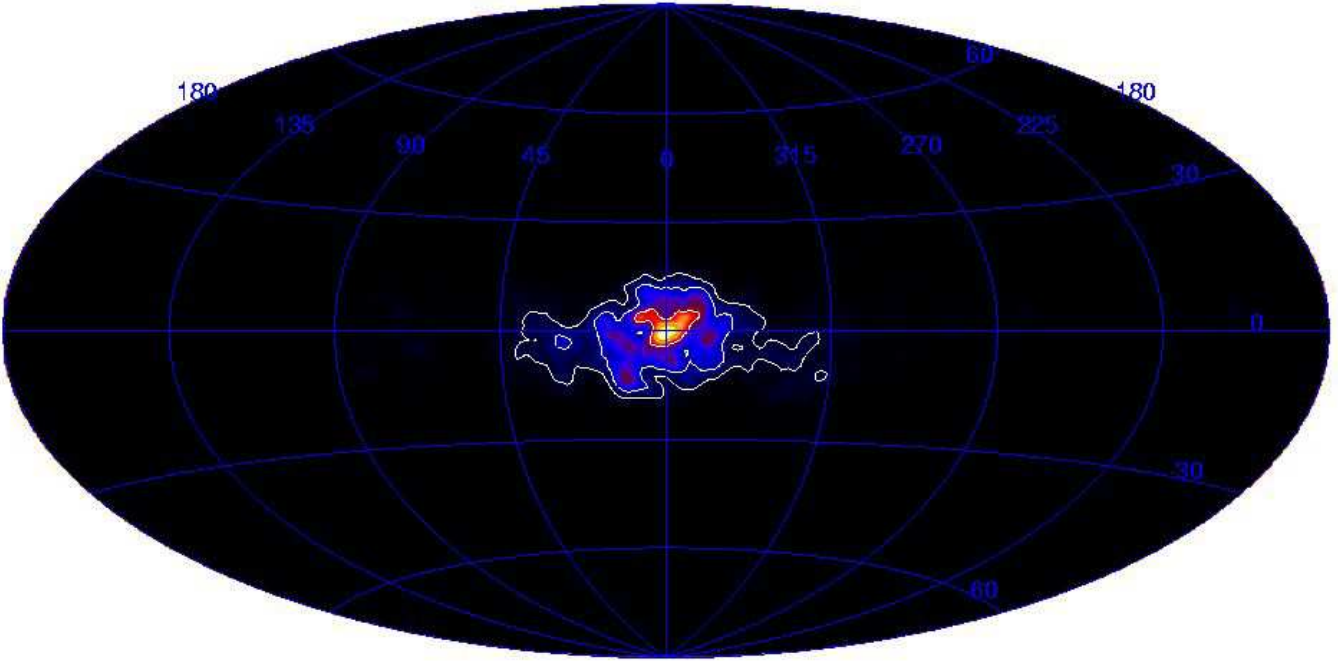}
  \caption{\label{fig:em25}
  Richardson-Lucy image after iteration 25.
  Contour levels are similar to Fig.~\ref{fig:em}.
  }
\end{figure*}

Figure~\ref{fig:em} reveals that the 511~keV sky is dominated by prominent 
emission from the bulge region of the Galaxy.
Beyond the galactic bulge, no additional 511~keV emission is seen all 
over the sky, despite the good exposure in some regions
(e.g.~Cygnus, Vela, LMC, anticentre, north galactic pole region).
The 511~keV emission appears symmetric and centred on the galactic centre, 
with indications for a slight latitude flattening.
The latitude flattening could be either due to an inherent asymmetry of 
the bulge component or due to the presence of an underlying faint galactic 
disk component.
Indeed, if the Richardson-Lucy iterations are continued, a faint 
disk-like structure emerges (c.f.~Fig.~\ref{fig:em25}).
Yet the image starts to become polluted by noise and we cannot exclude 
the possibility that the apparent disk emission is artificially created 
by the exposure pattern that follows the galactic plane.
Therefore we employ more quantitative methods in the next section to 
assess the significance of the possible disk emission.

By fitting Gaussian functions to the longitude and latitude profiles 
of the image (c.f.~Fig.~\ref{fig:emprofiles}) we estimate 
the extent of the emission to $\sim13\deg \times 10\deg$ (FWHM).
Figure \ref{fig:emprofiles} indicates, however, that the emission profiles 
are not well represented by Gaussian functions.
The emission is better described by a compact (FWHM $\sim5\deg$) core and 
a more extended halo (FWHM $10\deg-20\deg$).
We want to emphasise, however, that this qualitative analysis should 
not be pushed too far, since image deconvolution is a non-linear 
process which is easily affected by image noise and exposure biases.

\subsection{Morphological characterisation}
\label{sec:morphology}

\subsubsection{Method}

To make a quantitative assessment of the morphology of the 511 keV 
line emission we use a maximum likelihood multi-component model 
fitting algorithm.
Assuming Poisson noise for the measured number $n_i$ of events in each 
of the $N$ data-space bins, the algorithm maximises the log likelihood
\begin{equation}
 \ln L = \sum_{i=1}^N n_{i} \ln e_{i} - e_{i} - \ln n_{i}! ,
 \label{eq:likelihood}
\end{equation}
where $e_i = \sum_{k} \alpha_k s^k_i + b_i(\bgpars)$ is the 
predicted number of (source plus background) counts in data space bin $i$,
$s^k_i = \sum_{j=1}^M f^k_j R_{ij}$ is the sky intensity model $f^k_j$
folded into the data space ($R_{ij}$ being the instrumental response 
matrix),
$b_i(\bgpars)$ is the background model (c.f.~Fig.~\ref{eq:bgm}),
and $\alpha_k$ and $\bgpars$ are scaling factors for the sky intensity 
and the background model, respectively, that are adjusted by the fit.

Detection significances (and parameter errors) are estimated using the 
maximum likelihood ratio test (Cash~\cite{cash79}).
We calculate the maximum log likelihood-ratio
$\mlr = -2 ( \ln L_0 - \ln L_1 )$ 
between two models (hypotheses), where for the first one we constrain 
a number $p$ of the parameters to specific values (resulting in $L_0$) 
while for the second one all parameters are left free (resulting in $L_1$).
In the case that $L_1$ provides a satisfactory fit of the data, \mlr\ is 
then distributed like a $\chi^2_p$ distribution with $p$ degrees of 
freedom.
Statistical parameter errors were estimated using the formalism of 
Strong~(\cite{strong85}).
Throughout this paper the error bars quoted are $1\sigma$.

We call the maximum log likelihood-ratio (\mlr) of a model 
the difference between the log likelihood obtained by fitting all 
model parameters and the log likelihood obtained by fitting only the 
background model to the data
(i.e.~for $L_1$ all parameters $\alpha_k$ and $\bgpars$ vary freely while 
for $L_0$ all $\alpha_k$ are constrained to zero and only the $\bgpars$ 
are allowed to vary).
To compare models with different numbers of free parameters, we quote 
the reduced maximum log likelihood-ratio, $\rmlr = \mlr - \dof$, with
\dof\ being the number of free parameters $\alpha_k$ of the sky intensity 
model.

\subsubsection{2d surface brightness distribution}
\label{sec:gaussian}

As a first step we characterise the apparent morphology of the 511 keV 
line emission on the sky using a 
2d angular Gaussian surface brightness distribution
for which we determined
the centroid, $l_0, b_0$,
the longitude and latitude extent, $\Delta l, \Delta b$, and
the 511~keV line flux.
The results of this analysis are summarised in Table \ref{tab:gaussfit}, 
the best fitting model intensity distribution is shown in 
Fig.~\ref{fig:modelmaps}.

\begin{table}[!t]
  \footnotesize
  \caption{\label{tab:gaussfit}
  Morphology of the emission assuming a 2d angular Gaussian surface 
  brightness distribution.
  }
  \begin{flushleft}
	\begin{tabular}{lc}
	\hline
	\hline
	\noalign{\smallskip}
	Quantity & Measured Value \\ 
	\noalign{\smallskip}
	\hline
	\noalign{\smallskip}
	\rmlr\ (\dof)                   & $462.2$ $(5)$ \\
	$l_0$                           & $-0.6\deg \pm 0.3\deg$ \\
	$b_0$                           & $+0.1\deg \pm 0.3\deg$ \\
	$\Delta l$ (FWHM)               & $8.1\deg \pm 0.9\deg$ \\
	$\Delta b$ (FWHM)               & $7.2\deg \pm 0.9\deg$ \\
	511 keV flux ($10^{-3}$ \funit) & $1.09 \pm 0.04$ \\
	\noalign{\smallskip}
	\hline
	\end{tabular}
  \end{flushleft}
\end{table}

The analysis confirms our earlier findings
(Jean et al.~\cite{jean03a};
Kn\"odlseder et al.~\cite{knoedl04a}; 
Weidenspointner et al.~\cite{weidenspointner04})
of a compact and symmetric 511~keV line emission distribution towards 
the galactic centre.
The centroid of the emission appears slightly offset from the 
galactic centre direction, at the statistical $2\sigma$ level, but we 
do not claim that this offset is significant.
From our earlier analyses we learned that the centroid can be shifted 
by this amount simply from the combined effect of statistical and 
systematic biases in the modelling of the instrumental background.

Within the statistical uncertainties, the emission appears fully
symmetric, with an extension of $\sim8\deg$ (FWHM).
Formally, we determine a marginal emission flattening of
$\Delta b / \Delta l = 0.89 \pm 0.14$.
The total 511~keV flux is
$(1.09 \pm 0.04) \times 10^{-3}$ \funit, where the quoted error includes 
the uncertainty in the extent of the emission and the statistical 
and systematic measurement errors (c.f.~section \ref{sec:obs}).

The \rmlr\ of $462.2$ that has been obtained using the \orbit\ 
background model converts into a formal detection significance of 
$22\sigma$.
Using the \dete\ background model and including the systematic 
uncertainties results in a substantially higher detection significance 
of $34\sigma$.
Neglecting systematic uncertainties would even boost the detection 
significance towards $49\sigma$.

\subsubsection{Galactic models}
\label{sec:galactic}

\begin{table*}[!t]
  \footnotesize
  \caption{\label{tab:modelfit}
  Galaxy model fitting results (see text).
  The columns give
  (1) the model,
  (2) the \rmlr\ (obtained using the \orbit\ background model) 
      and the number of free model parameters (DOF),
  (3) the bulge scale length,
  (4) the bulge scale height,
  (5)-(8) the 511~keV photon luminosity of the model components, and
  (9)-(12) the 
  total $4\pi$ integrated 511~keV line all-sky flux
  in each of the model components.
  The prime indicates model components for which the scaling 
  parameters were adjusted by the fit.
  The figures in parenthesis quoted in columns (5)-(12) indicate $1\sigma$ 
  uncertainties in the last digit.
  }
  \begin{flushleft}
  \begin{tabular}{lccccccc@{\extracolsep{0.9em}}cccc}
    \hline
    \hline
    \noalign{\smallskip}
    Model & \rmlr\ & $R_0$ & $z_0$ 
    & \multicolumn{4}{c}{$\lgamma$ ($10^{43}$ \emunit)} 
    & \multicolumn{4}{c}{511 keV line flux ($10^{-3}$ \funit)} \\
    \cline{5-8}
    \cline{9-12}
    & (DOF) & (kpc) & (kpc) 
    & bulge & disk & halo & total 
    & bulge & disk & halo & total \\ 
    \noalign{\smallskip}
    \hline
    \noalign{\smallskip}
    G0 & 447.5 (1) & $0.91$ & $0.51$
      & $1.04  (3)$ &             &             & $1.04  (3)$
      & $1.22  (4)$ &             &             & $1.22  (4)$ \\
    G1 & 445.7 (1) & & 
      & $0.97  (3)$ &             &             & $0.97  (3)$
      & $1.19  (4)$ &             &             & $1.19  (4)$ \\
    G2 & 450.8 (1) & & 
      & $0.98  (3)$ &             &             & $0.98  (3)$
      & $1.18  (3)$ &             &             & $1.18  (3)$ \\
    G3 & 462.2 (1) & & 
      & $0.98  (3)$ &             &             & $0.98  (3)$
      & $1.18  (3)$ &             &             & $1.18  (3)$ \\
    E1 & 441.8 (1) & &
      & $0.99  (4)$ &             &             & $0.99  (4)$
      & $1.19  (4)$ &             &             & $1.19  (4)$ \\
    E2 & 453.1 (1) & &
      & $1.01  (3)$ &             &             & $1.01  (3)$
      & $1.19  (3)$ &             &             & $1.19  (3)$ \\
    E3 & 464.9 (1) & &
      & $1.00  (3)$ &             &             & $1.00  (3)$
      & $1.17  (3)$ &             &             & $1.17  (3)$ \\
    S$_{\rm F}$ & 459.0 (1) & &
      & $0.96  (3)$ &             &             & $0.96  (3)$
      & $1.19  (3)$ &             &             & $1.19  (3)$ \\
    E$_{\rm F}$ & 459.0 (1) & &
      & $0.93  (2)$ &             &             & $0.93  (2)$
      & $1.19  (3)$ &             &             & $1.19  (3)$ \\
    P$_{\rm F}$ & 456.5 (1) & &
      & $0.94  (3)$ &             &             & $0.94  (3)$
      & $1.16  (3)$ &             &             & $1.16  (3)$ \\
    S$_{\rm PR}$ & 456.3 (1) & &
      & $0.94  (2)$ &             &             & $0.94  (2)$
      & $1.22  (3)$ &             &             & $1.22  (3)$ \\
    \noalign{\smallskip}
    \hline
    \noalign{\smallskip}
    G0' & 462.5 (3) & $0.52 (6)$ & $0.45 (5)$
      & $0.94  (4)$ &             &             & $0.94  (4)$
      & $1.09  (4)$ &             &             & $1.09  (4)$ \\
    E0' & 464.2 (3) & $0.37 (5)$ & $0.42 (7)$
      & $0.98  (5)$ &             &             & $0.98  (5)$
      & $1.15  (5)$ &             &             & $1.15  (5)$ \\
    H' & 468.4 (4) & & 
      &             &             & $1.6   (3)$ & $1.6   (3)$
      &             &             & $2.2   (4)$ & $2.2   (4)$ \\
    Shells & 469.0 (2) & & 
      & $0.97  (3)$ &             &             & $0.97  (3)$
      & $1.13  (3)$ &             &             & $1.13  (3)$ \\
    \noalign{\smallskip}
    \hline
    \noalign{\smallskip}
    E3+D0 & 466.3 (2) & &
      & $0.95  (3)$ & $0.11  (5)$ &             & $1.05  (4)$
      & $1.11  (4)$ & $0.4   (2)$ &             & $1.53  (5)$ \\
    G0'+D0 & 465.2 (4) & $0.48 (6)$ & $0.46 (6)$
      & $0.87  (4)$ & $0.15  (5)$ &             & $1.03  (5)$
      & $1.01  (5)$ & $0.6   (2)$ &             & $1.61  (7)$ \\
    E0'+D0 & 466.2 (4) & $0.34 (4)$ & $0.44 (8)$
      & $0.92  (5)$ & $0.14  (5)$ &             & $1.06  (5)$
      & $1.07  (6)$ & $0.5   (2)$ &             & $1.61  (8)$ \\
    H'+D0 & 468.2 (5) & &
      &             & $0.09  (8)$ & $1.2   (3)$ & $1.3   (3)$
      &             & $0.4   (3)$ & $1.6   (5)$ & $2.1   (5)$ \\
    Shells+D0 & 472.2 (3) & &
      & $0.91  (4)$ & $0.15  (5)$ &             & $1.05  (4)$
      & $1.05  (4)$ & $0.6   (2)$ &             & $1.62  (6)$ \\
    \noalign{\smallskip}
    \hline
    \noalign{\smallskip}
    E3+D1 & 468.8 (2) & &
      & $0.93  (4)$ & $0.23  (8)$ &             & $1.15  (6)$
      & $1.09  (5)$ & $0.8   (3)$ &             & $1.90  (9)$ \\
    G0'+D1 & 468.6 (4) & $0.47 (6)$ & $0.45 (6)$
      & $0.84  (5)$ & $0.31  (9)$ &             & $1.15  (6)$
      & $0.98  (5)$ & $1.1   (3)$ &             & $2.1   (1)$ \\
    E0'+D1 & 469.5 (4) & $0.33 (4)$ & $0.42 (8)$
      & $0.89  (5)$ & $0.29  (9)$ &             & $1.17  (7)$
      & $1.03  (6)$ & $1.0   (3)$ &             & $2.1   (1)$ \\
    H'+D1 & 470.4 (5) & &
      &             & $0.3   (1)$ & $1.2   (3)$ & $1.4   (3)$
      &             & $0.9   (4)$ & $1.5   (4)$ & $2.4   (5)$ \\
    Shells+D1 & 474.9 (3) & & 
      & $0.88  (4)$ & $0.29  (8)$ &             & $1.17  (6)$
      & $1.03  (5)$ & $1.0   (3)$ &             & $2.0   (1)$ \\
    \noalign{\smallskip}
    \hline
    \end{tabular}
  \end{flushleft}
\end{table*}

To determine the galactic positron-electron annihilation rate requires 
modelling the spatial distribution of the positron-electron 
annihilation. 
The 511 keV photon luminosity \lgamma\ is related to the positron 
luminosity \lpositron\ through 
$\lgamma = (2 - 1.5 \fpositron) \times \lpositron$
where \fpositron\ is the positronium (Ps) fraction, defined as the 
fraction of positrons that decay via positronium formation
(Brown, \& Leventhal \cite{brown87}).
Using $\fpositron=0.93 \pm 0.04$ that has been determined from OSSE 
observations (Kinzer et al.~\cite{kinzer01})
results in a conversion from 511 keV photon luminosity to 
a positron-electron annihilation rate of
$\lpositron = (1.64 \pm 0.06) \times \lgamma$.

We here compare models of bulge, disk, and halo components with the 
data.
Based on galactic model density distributions $\rho(x,y,z)$ we calculate 
the expected all-sky 511 keV intensity $f(l,b)$ towards direction $(l,b)$ 
by integrating the volume emissivity $\rho(x,y,z)$ along the line of sight 
$s$:
\begin{equation}
f(l,b) = \frac{1}{4\pi} \int \rho(x,y,z) {\rm d}s
\end{equation}
(the galactic centre has been assumed to be at a distance of 
$\Rsol = 8.5$ kpc).
Galactic 511~keV photon luminosities are calculated by integrating 
$\rho(x,y,z)$ over the galactic volume, 
\begin{equation}
\lpositron = \int \rho(x,y,z) s^2 {\rm d}s {\rm d}\Omega
\end{equation}
assuming an outer Galaxy radius of $R_{\rm max}=15$ kpc.

\begin{figure*}[!t]
  \center
  \epsfxsize=8.8cm \epsfclipon
  \epsfbox{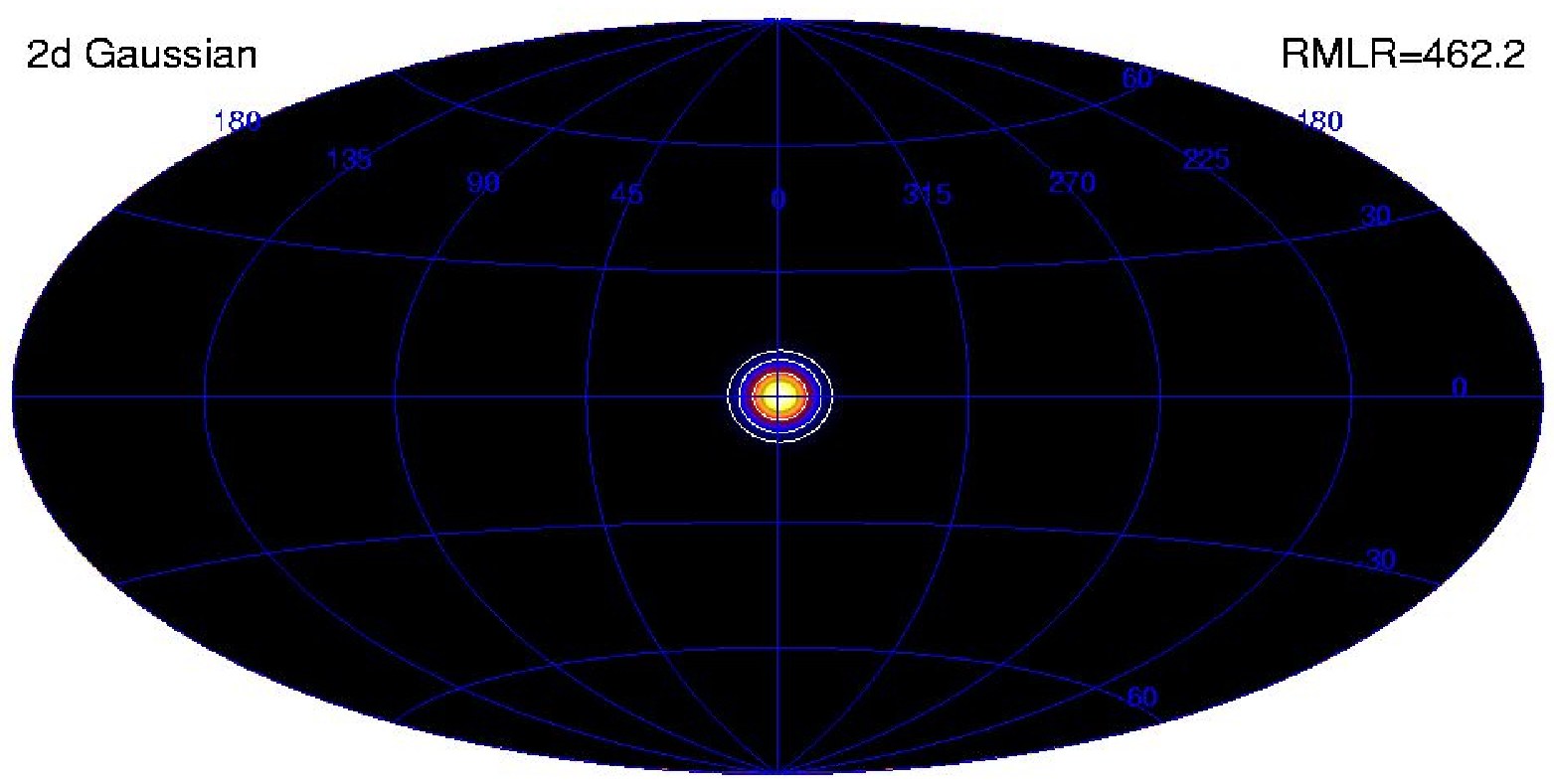}
  \hfill
  \epsfxsize=8.8cm \epsfclipon
  \epsfbox{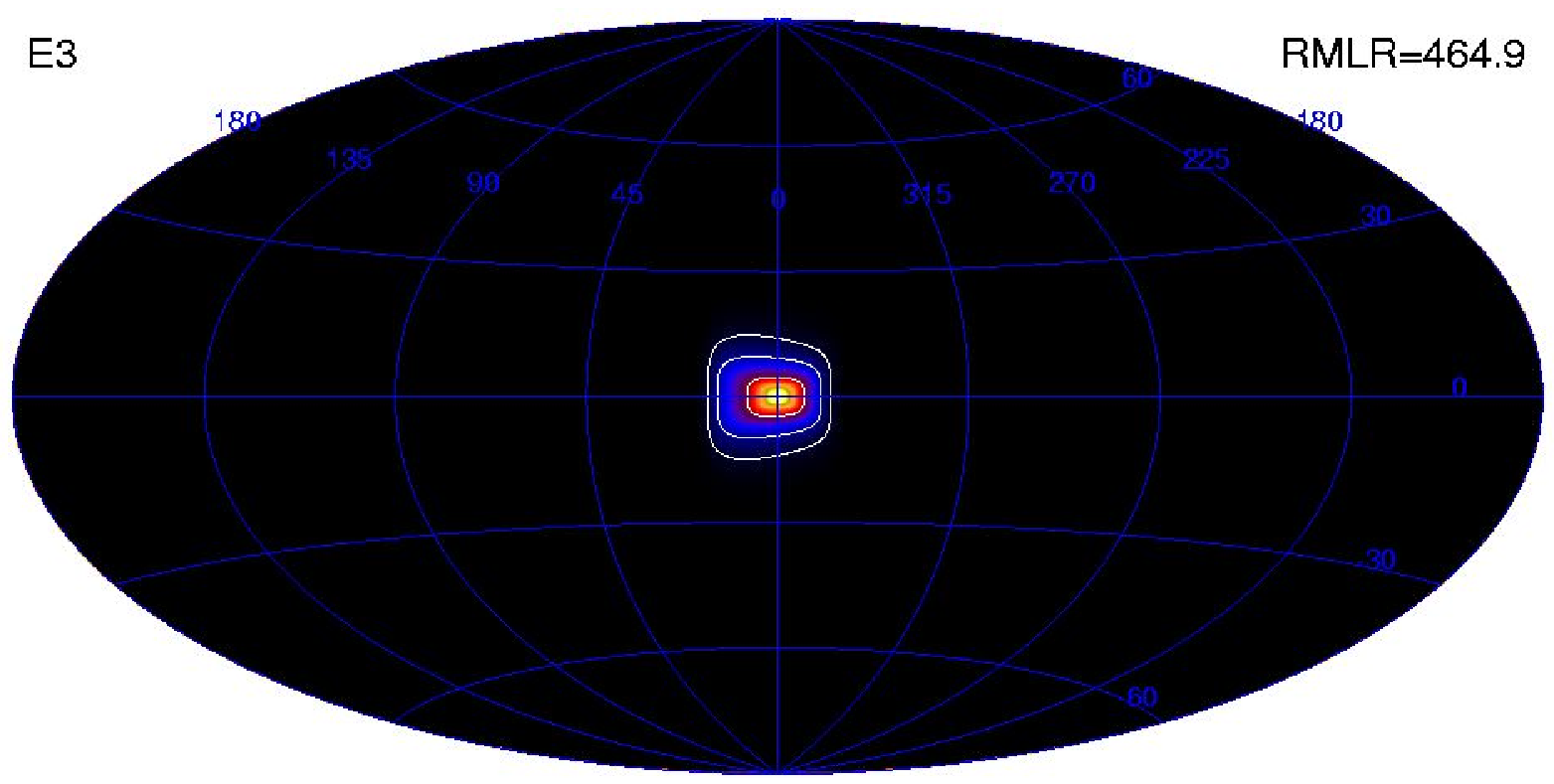}
  \epsfxsize=8.8cm \epsfclipon
  \epsfbox{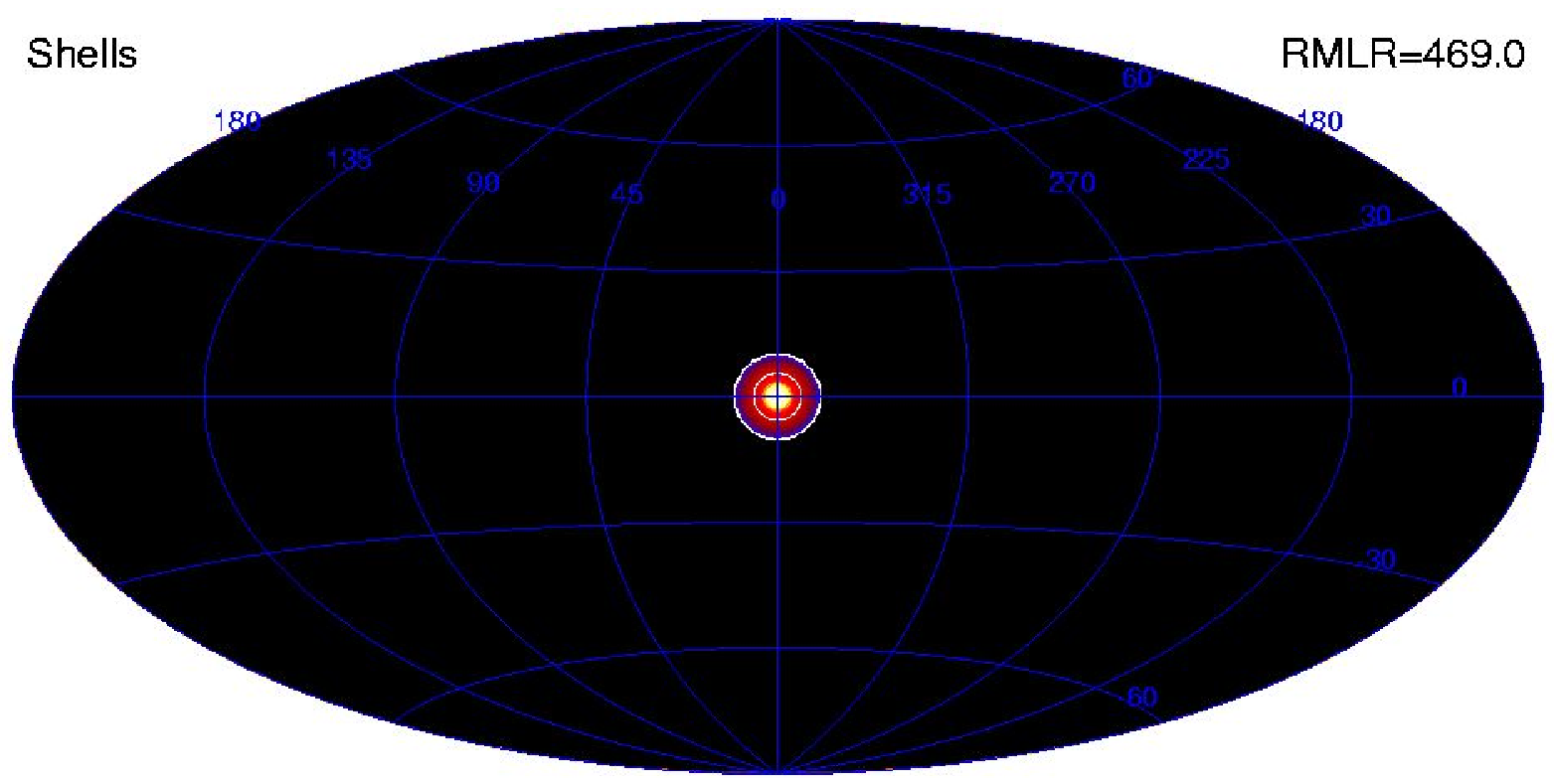}
  \hfill
  \epsfxsize=8.8cm \epsfclipon
  \epsfbox{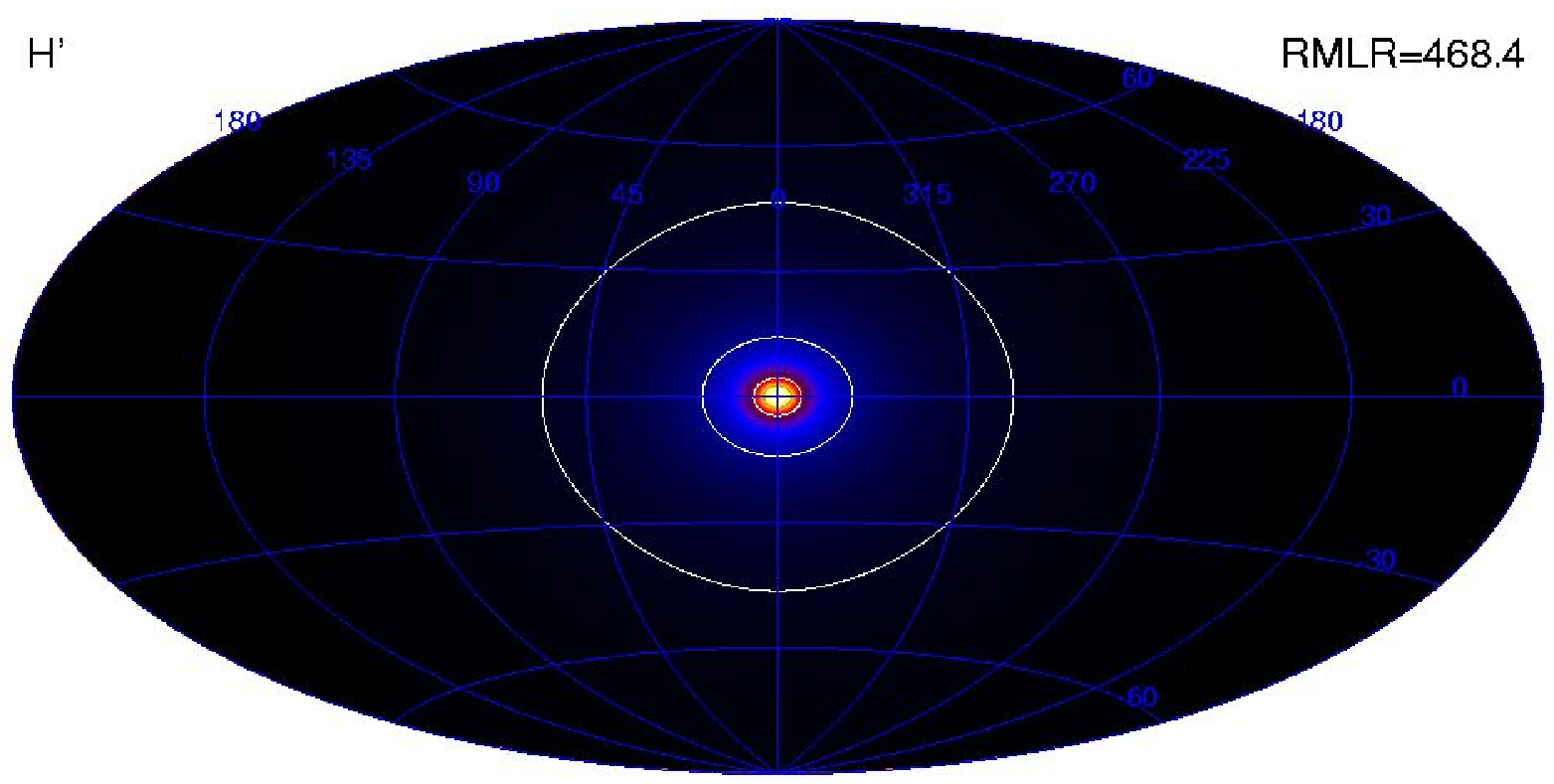}
  \epsfxsize=8.8cm \epsfclipon
  \epsfbox{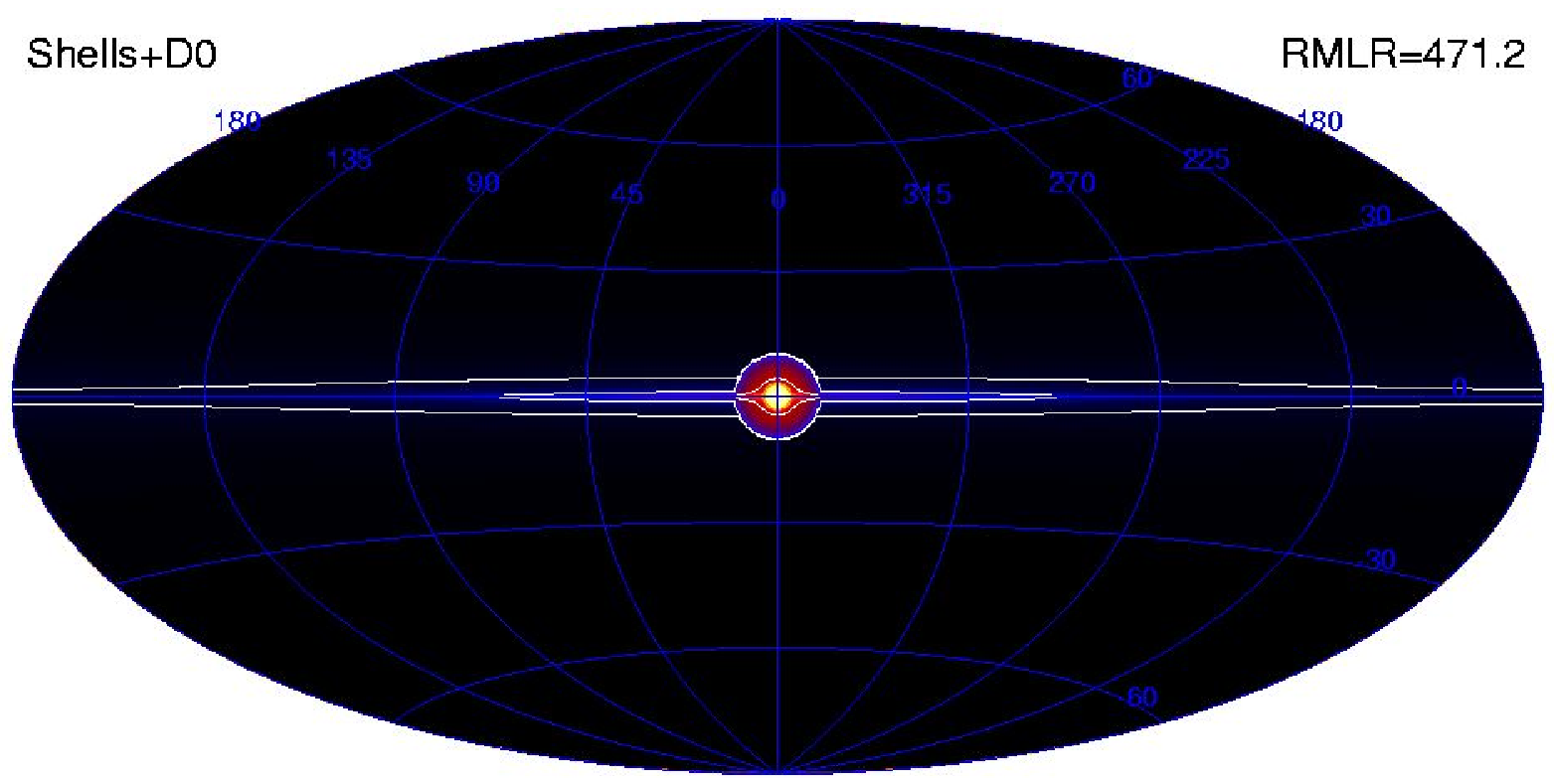}
  \hfill
  \epsfxsize=8.8cm \epsfclipon
  \epsfbox{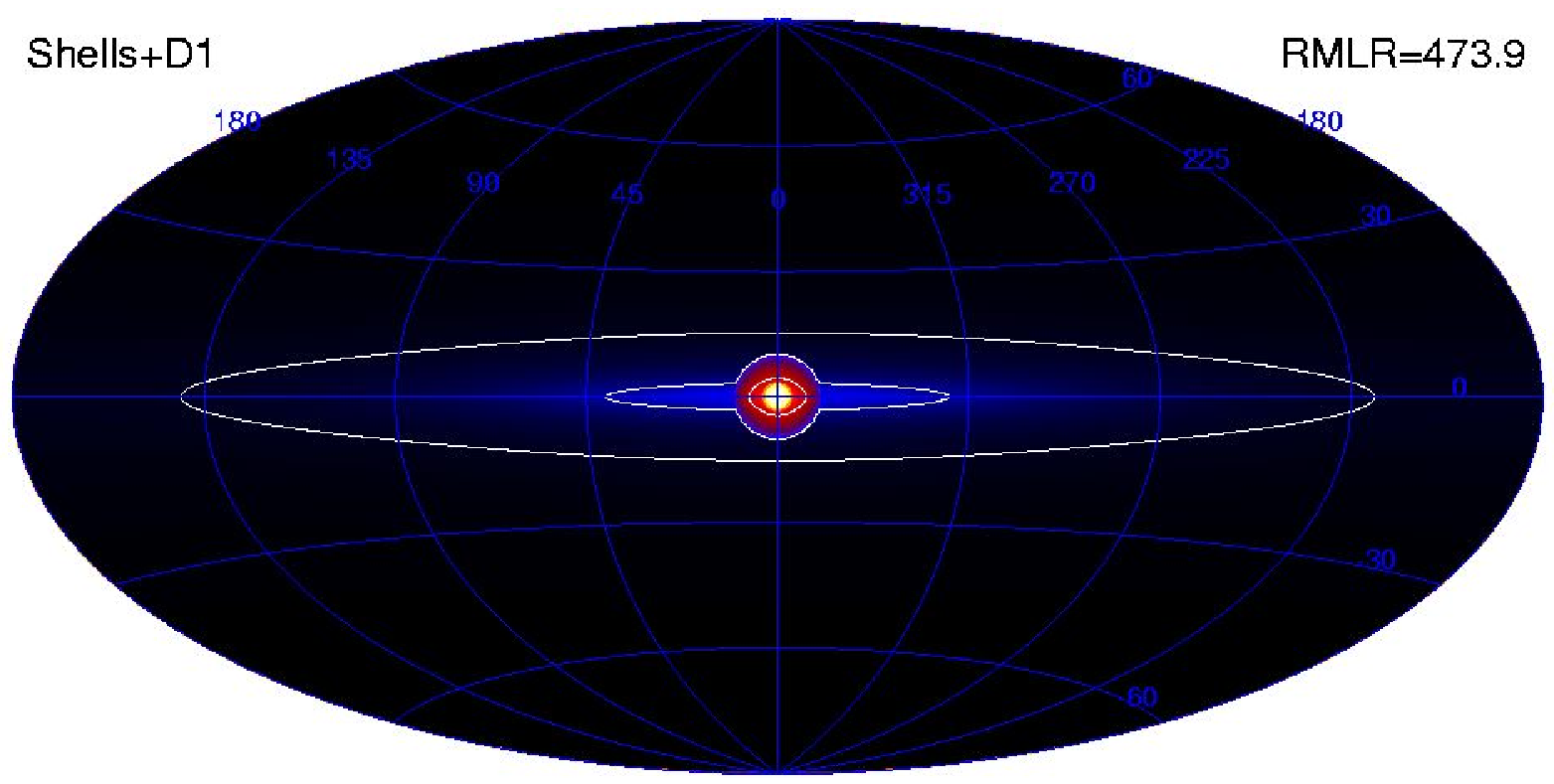}
  \epsfxsize=8.8cm \epsfclipon
  \epsfbox{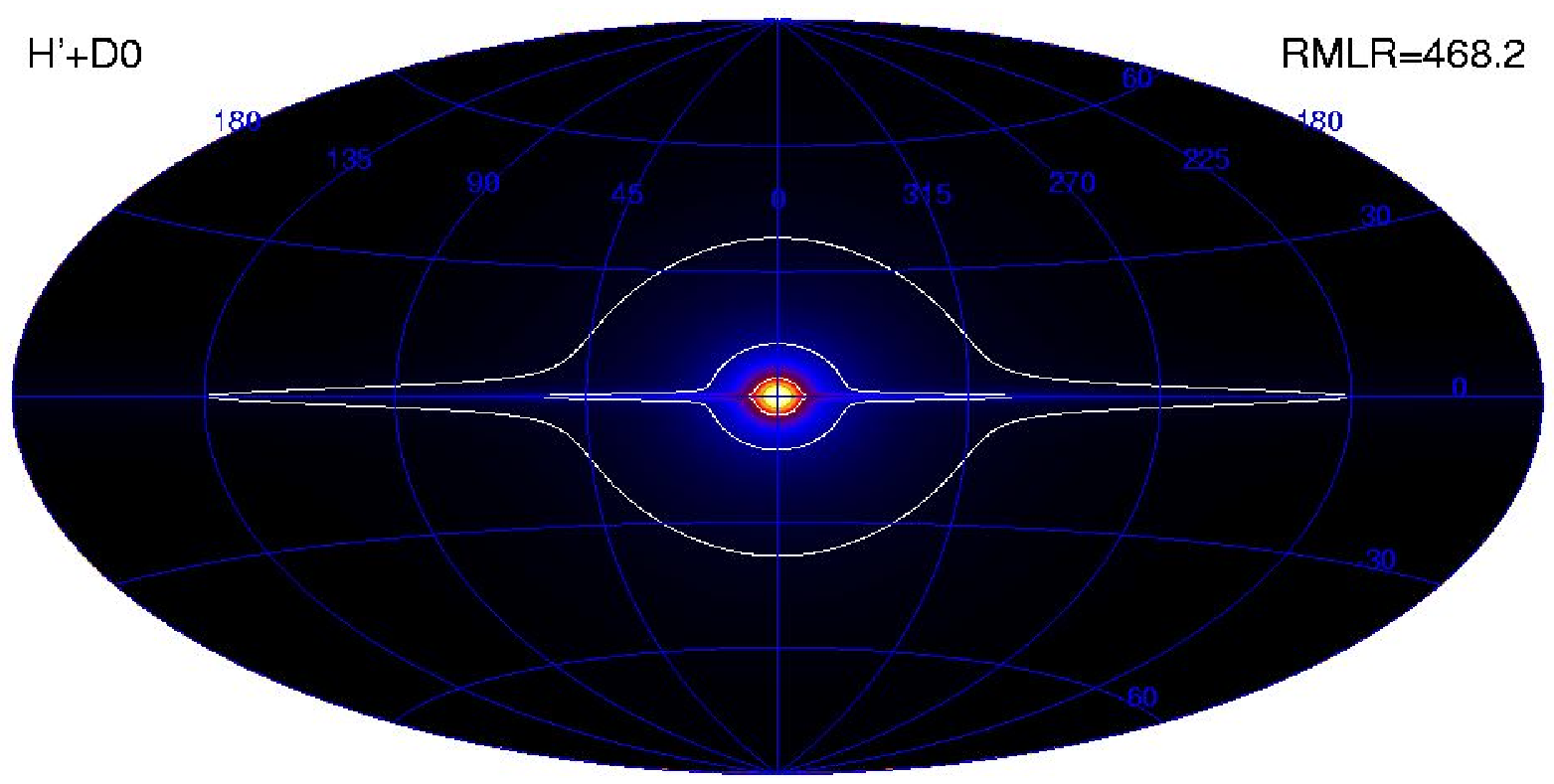}
  \hfill
  \epsfxsize=8.8cm \epsfclipon
  \epsfbox{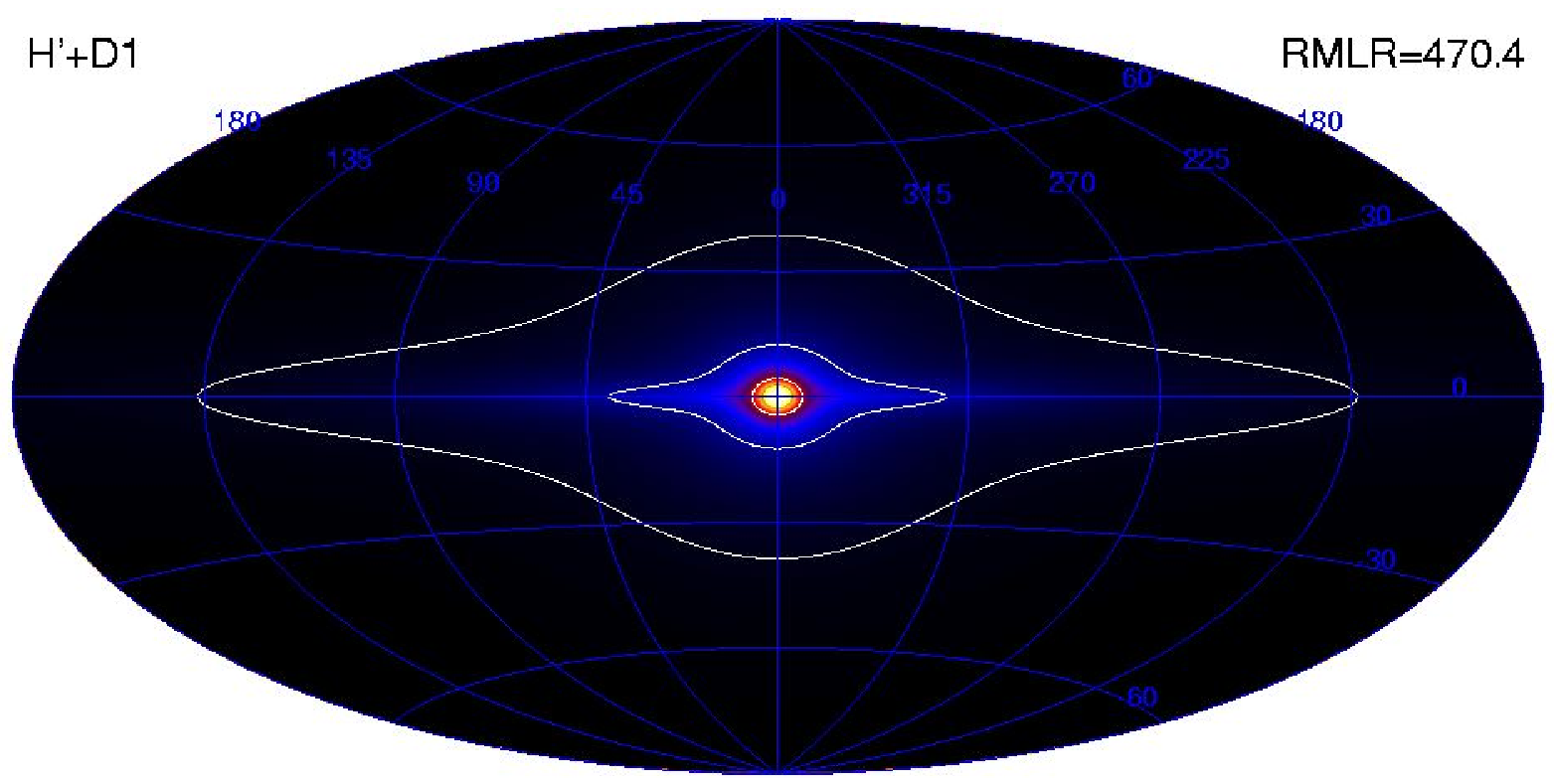}
  \caption{\label{fig:modelmaps}
  All-sky maps of the best fitting models of 511 keV gamma-ray line 
  emission (see text for a description of the models).
  Contour levels indicate intensity levels (from the centre outwards)
  of $10^{-2}$, $10^{-3}$, and $10^{-4}$ \fster.
  The resulting \rmlr s of the model fits are quoted in the upper-right 
  corner of the panels.
  }
\end{figure*}

Since the 511~keV line emission is primarily arising from the galactic 
centre region we fitted in a first step models of the galactic stellar bulge 
to the data.
To account for uncertainties in our knowledge about the morphology of 
this component (which are related to our location in the galactic 
plane amid the obscuration by interstellar dust) we compared a variety of 
proposed bulge models to the data.
The models were gathered from
Dwek et al.~(\cite{dwek95}) and Freudenreich (\cite{freudenreich98}),
who modelled the distribution of K and M giant stars using DIRBE near-infrared 
skymaps, and from Picaud \& Robin (\cite{picaud04}) who analysed data 
from the DENIS near-infrared survey.
There is an accumulating body of evidence that the stellar 
distribution in the bulge is bar-shaped, and except for models G0 and 
E0, all employed bulge models have triaxial morphologies that differ in the 
orientation angles, the scale lengths, and the radial density profiles.
Details of the models are given in Appendix \ref{sec:gal-models},
the results of the analysis are summarised in Table 
\ref{tab:modelfit}, and best fitting 511~keV intensity distributions 
are shown in Fig.~\ref{fig:modelmaps}.

The best fitting bulge models are E3, G3, S$_{\rm F}$, and E$_{\rm F}$.
Reasonably good fits are also obtained for P$_{\rm F}$ and S$_{\rm PR}$, 
while only moderate fits are achieved for the remaining models.
Our ranking is similar to that established from the analysis of the 
DIRBE and DENIS near-infrared data
(Dwek et al.~\cite{dwek95}; Freudenreich \cite{freudenreich98}; 
Picaud \& Robin \cite{picaud04}).
The best fitting bulge models fit the data as well as the adjusted
2d angular Gaussian surface brightness distribution.
This means that models of the galactic stellar bulge are able to 
explain satisfactorily the morphology of the 511~keV bulge emission.

In a second step we fitted the 511~keV emission using parametric models 
of the galactic bulge and halo morphology in order to determine the 
scale of the emission.
For the bulge models G0' and E0' we adjust the
radial scale length ($R_0$) and 
vertical scale height ($z_0$), 
while for the galactic halo model H' we determine the
density slope powerlaw index ($n$),
the inner cutoff radius ($a_{\rm c}$), and
the axis ratio ($\epsilon$).
In addition, we employed a model composed of a set of galactocentric 
nested shells of constant density (model `Shells') to determine the radial 
density profile of the 511~keV emission.
We varied the radii of the shells and the number of shells in order to 
maximise the \mlr, whilst limiting the number of shells to the minimum 
required to satisfactorily describe the data.

\begin{figure}[!t]
  \center
  \epsfxsize=8.8cm \epsfclipon
  \epsfbox{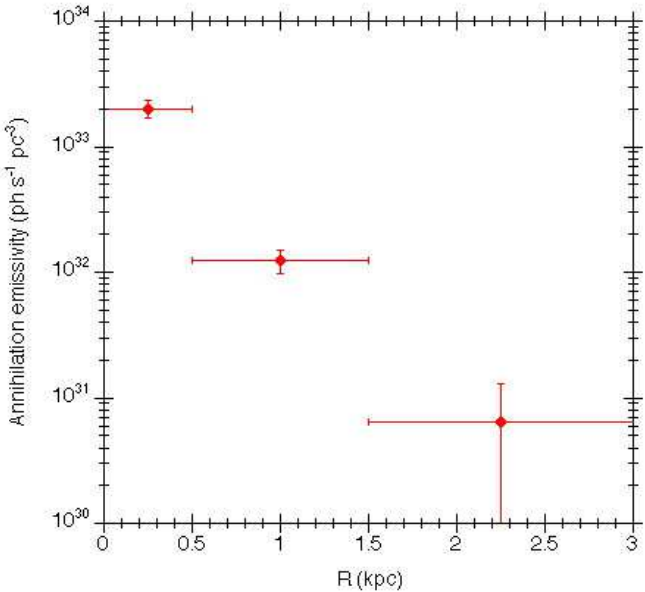}
  \caption{\label{fig:radial}
  Radial dependence of the 511~keV volume emissivity as derived from 
  the bulge model `Shells'.
  }
\end{figure}

The data suggest a symmetric bulge emission profile, with scale lengths 
between $300$ and $600$ pc.
The \rmlr s are comparable to the best fitting bulge models that we tested 
before.
The data are equally well fitted by a model of the galactic halo, with a 
density powerlaw index of $n=3.0 \pm 0.3$,
an inner cutoff radius of $a_{\rm c}=0.39 \pm 0.08$ kpc, and 
a flattening of $\epsilon = 0.81 \pm 0.12$.
Most studies of the stellar halo population suggest power indices between
$2.4$ and $3.5$ and flattenings in the range $0.6$ to $1.0$, while the 
inner cutoff radius is basically undetermined
(Robin et al.~\cite{robin00} and references therein).
Our values are compatible with those of the stellar halo population, 
but the large uncertainties in the stellar halo morphology do not 
allow firm conclusions to be drawn.

The nested shell model provides the best fit to the data thanks to 
its flexibility in adjusting the radial density profile of the emission.
A satisfactory fit is achieved by using two shells with radii 
$0-0.5$ and $0.5-1.5$ kpc; splitting up these shells in a finer 
binning, moving the shell interface radius or adding more shells does 
not significantly improve the fit.
In particular, we detect no significant 511~keV bulge emission from 
galactocentric distances $\ga 1.5$ kpc.
The radial dependence of the 511~keV volume emissivity is plotted in
Fig.~\ref{fig:radial}.
For illustration we added the result of a third shell to the figure that 
covers radial distances of $1.5-3.0$ kpc and for which the flux is 
consistent with zero.
Our fit reveals a drop in the annihilation emissivity by one 
order of magnitude between the inner $0-0.5$~kpc and the outer 
$0.5-1.5$~kpc shell, confirming the existence of a narrow core plus an 
extended halo of 511~keV emission that has already been suggested by 
the imaging analysis (c.f.~section \ref{sec:map}).

In a third step we added galactic disk components to the bulge and 
halo models.
For the galactic disk we tested models of young (model D0) and old (model D1) 
stellar populations (Robin et al.~\cite{robin03}).
With both models we find clear evidence for 511~keV line emission 
from the galactic disk.
Adding disk models D0 and D1 to bulge or halo models consistently 
improves the fit leading to a detection of the disk emission at the
$3-4\sigma$ level.\footnote{
  Using the quoted \rmlr s, the formal significance of the disk 
  emission amounts only to $2-3\sigma$.
  However, the tabulated \rmlr s have been obtained using the \orbit\ 
  background model which is less sensitive to 511~keV line emission
  than the \dete\ model.
  Using the procedure outlined in section \ref{sec:bgm} we reduce the 
  flux uncertainties and increase the detection significance to
  $3-4\sigma$.}
Formally, D1 provides a better fit than D0, but the difference is 
marginal.
The signal from the disk is still too faint in our present dataset to 
deduce anything about its morphology.

\begin{table}[!t]
  \footnotesize
  \caption{\label{tab:fitsummary}
  Summary of model fitting results.
  Fluxes are given as total $4\pi$ integrated all-sky values.
  Annihilation rates have been calculated assuming $\fpositron=0.93$.
  }
  \begin{flushleft}
  \begin{tabular}{l@{\extracolsep{0.4em}}c@{\extracolsep{1em}}c@{\extracolsep{1em}}c}
  \hline
  \hline
  \noalign{\smallskip}
  Quantity & Bulge & Halo & Disk \\ 
  \noalign{\smallskip}
  \hline
  \noalign{\smallskip}
  Flux ($10^{-3}$ \funit) & 
    $1.05 \pm 0.06$ & $1.6 \pm 0.5$ & $0.7 \pm 0.4$ \\
  $\lgamma$ ($10^{43}$ \emunit) &
    $0.90 \pm 0.06$ & $1.2 \pm 0.3$ & $0.2 \pm 0.1$ \\
  $\lpositron$ ($10^{43}$ \peunit) &
    $1.50 \pm 0.10$ & $2.0 \pm 0.5$ & $0.3 \pm 0.2$ \\
  \noalign{\smallskip}
  \hline
  \noalign{\smallskip}
  \multicolumn{3}{l}{Total flux ($10^{-3}$ \funit)}          & $1.5-2.9$ \\
  \multicolumn{3}{l}{Total $\lgamma$ ($10^{43}$ \emunit)}    & $1.0-1.7$ \\
  \multicolumn{3}{l}{Total $\lpositron$ ($10^{43}$ \peunit)} & $1.6-2.8$ \\
  \multicolumn{3}{l}{B/D flux ratio}                         & $1-3$ \\
  \multicolumn{3}{l}{B/D luminosity ratio}                   & $3-9$ \\
  \noalign{\smallskip}
  \hline
  \end{tabular}
  \end{flushleft}
\end{table}

The flux, luminosity and annihilation rate in the bulge, halo and disk 
components are summarised in Table \ref{tab:fitsummary}.
Recall that the bulge and halo components are alternatives and 
their contribution should not be added to derive the total galactic 
values.
Either component provides an almost equally good fit to the data.
Due to their degeneracy fitting both simultaneously is not meaningful.

The halo model leads to a considerably larger flux, luminosity and 
annihilation rate than the bulge model due to the presence of a flat 
and extended tail in this distribution (c.f.~Fig.~\ref{fig:modelmaps}).
Currently, our data do not allow to detect this tail, and thus, they do not 
allow to discriminate between bulge and halo models.
Future deep observations at intermediate galactic latitudes that are 
scheduled for the INTEGRAL AO-3 observing period aim in measuring 
this emission tail, promising to provide constraints that will allow 
in the future to disentangle between the different emission morphologies.

The data suggest bulge-to-disk 511~keV flux ratios in the range
$1 - 3$, where the lower boundary is obtained for the short 
scale-length old stellar disk model D1 which suggests larger disk flux 
values than the young stellar disk model D0.
Halo-to-disk 511~keV flux ratios are even larger, in the range $2 - 
4$, owing to the larger flux in the halo component.
The large uncertainty in these ratios arises from the low 
intensity of the galactic disk component, which for the analysed 
dataset is just above the SPI detection limit.

We also note that the bulge-to-disk 511~keV photon luminosity ratio is 
much higher than the bulge-to-disk flux ratio and lies in the range 
$3 - 9$.
This difference is explained by the fact that the average squared 
distance 
$\bar{s}^2 = \int \rho(s) s^2 {\rm d}s {\rm d}\Omega / 
	     \int \rho(s) {\rm d}s {\rm d}\Omega$,
which defines the distance at which a source of luminosity \lpositron\
produces the observed 511~keV line flux, is smaller for the galactic 
disk than for the galactic bulge.
In other words, to produce the same 511~keV flux at Earth, the 
intrinsic luminosity of the bulge has to be larger than that of the 
disk.\footnote{
  The difference between bulge-to-disk flux and luminosity ratio is 
  only important for our home Galaxy and is related to the fact that 
  the Sun is located within the galactic radius.
  For external galaxies this difference disappears since their bulge and 
  disk appear at the same distance to us.}
It is therefore important to quote explicitly the quantity for which 
we discuss the bulge-to-disk ratio.
The same rational also holds for the halo-to-disk 511~keV photon luminosity 
ratio, which is larger than the corresponding flux ratio, and which 
is comprised between $4 - 13$.

\subsection{Correlation with tracer maps}
\label{sec:correlation}

\begin{figure}[!t]
  \center
  \epsfxsize=8.8cm \epsfclipon
  \epsfbox{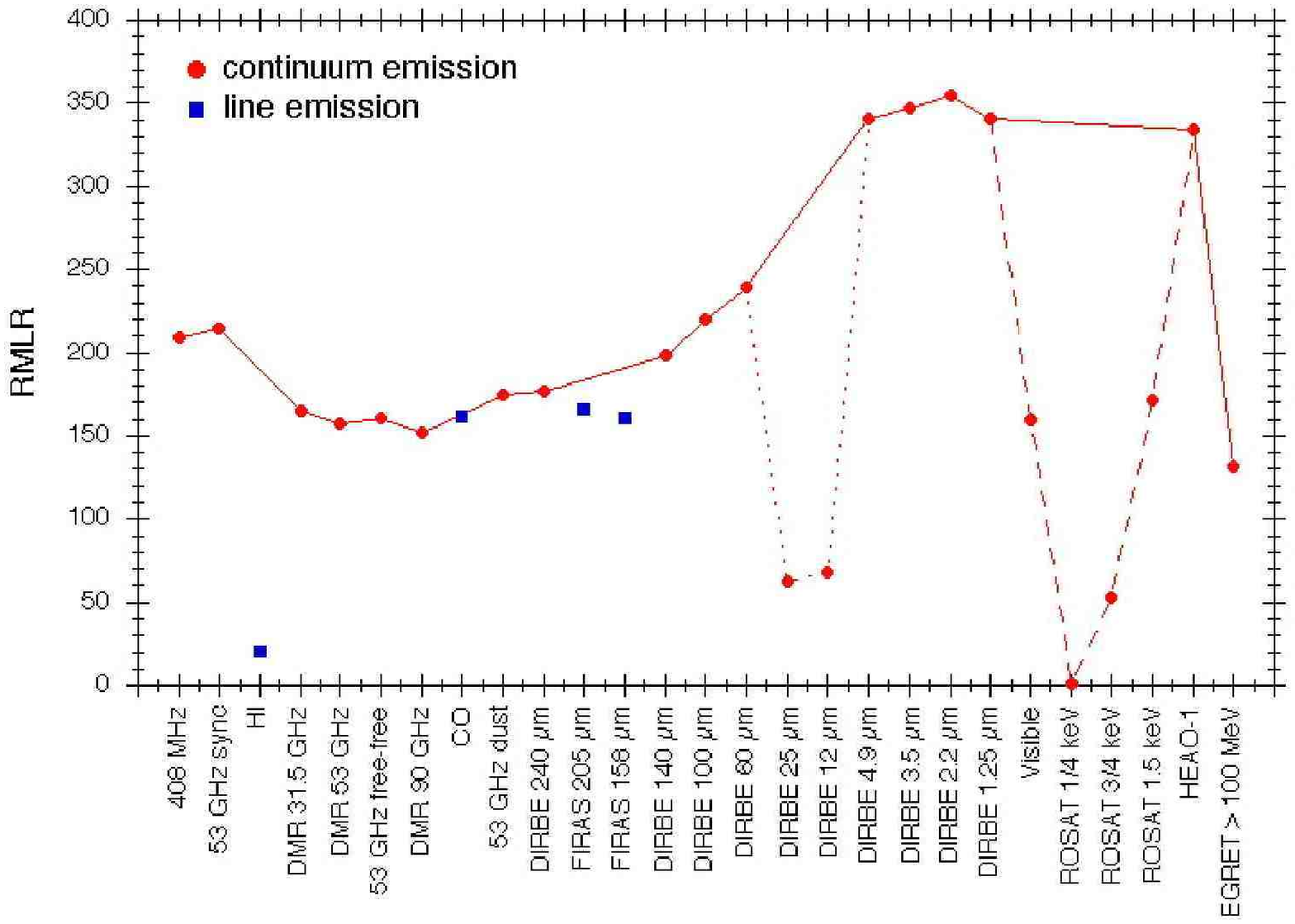}
  \caption{\label{fig:mlr-tracer}
  Reduced maximum log likelihood-ratio (\rmlr) as function of the tracer map, 
  ordered by increasing photon energy (or decreasing wavelength).
  The dotted line illustrates the impact of zodiacal light 
  contamination in the DIRBE 12 \um\ and 25 \um\ skymaps, the dashed 
  line indicates maps that are heavily affected by galactic absorption.
  }
\end{figure}

To gain insight into the nature of the galactic positron sources, we 
searched for correlations between the 511~keV line emission morphology 
and all-sky intensity distributions observed at other 
wavelengths.
This work was inspired by a similar study that
Kn\"odlseder et al.~(\cite{knoedl99b})
performed to understand the morphology of the 1.8 \MeV\ gamma-ray 
line emission (arising from radioactive decay of \al) observed by the 
COMPTEL telescope aboard CGRO.
Through their analysis, the authors could establish a tight 
correlation between the morphology of galactic microwave free-free 
emission and that of 1.8 \MeV\ line emission, hinting towards a 
massive star origin of \al.
The tracer maps used for the comparison are those listed in 
Kn\"odlseder et al.~(\cite{knoedl99b}).
For a detailed description of the maps the reader is referred to that 
work.

Figure~\ref{fig:mlr-tracer} summarises the result of the correlation 
study for our 511 keV dataset, where we show the \rmlr\ as a function of 
the tracer map (ordered by increasing photon energy or decreasing 
wavelength).
None of the tracer maps is consistent with the data.
The maximum \rmlr\ that is reached ($353.7$ for the DIRBE 2.2 \um\ 
map) is more than $100$ units smaller than the values obtained for the 
parametric models of the previous sections.
Apparently, the 511~keV emission morphology is unique and cannot be 
represented by any known celestial intensity distribution.

Nevertheless, Fig.~\ref{fig:mlr-tracer} shows a clear trend, where the 
data favour maps in the near-infrared domain
(DIRBE 1.25 \um\ - 4.9 \um)
and the hard X-ray band
(HEAO-1)
over maps observed at longer wavelengths.
In particular, the worst fits are obtained in the microwave and 
far-infrared domain where the skymaps trace the young stellar 
population, either through their ionising radiation
(DMR maps at $\nu \ga 53$ GHz),
or through their related molecular gas
(CO)
and cold dust emission
(DIRBE 100 \um\ - 240 \um).
From this it is clear that the bulk of the 511~keV emission is not 
related to the young massive stellar population of the Galaxy.

\begin{figure}[!t]
  \center
  \epsfxsize=8.8cm \epsfclipon
  \epsfbox{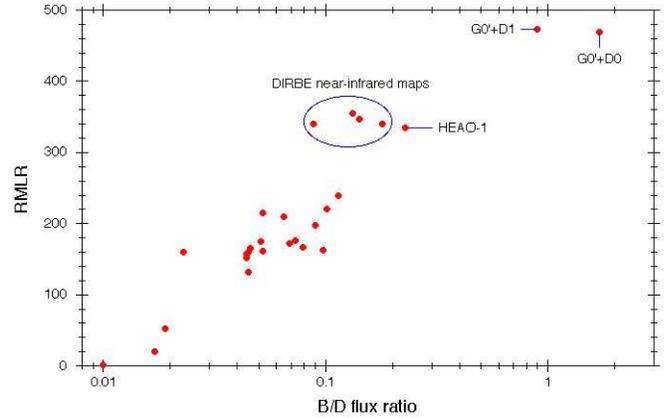}
  \caption{\label{fig:mlr-bd}
  Reduced maximum log likelihood-ratio (\rmlr) as function of the bulge-to-disk 
  flux ratio of the tracer map.
  The results for the parametric bulge+disk models are 
  also shown.
  }
\end{figure}

On the contrary, all best fitting tracers maps show the characteristic 
features of an old stellar population:
a strong bulge component combined with a short scale radius disk component.
Apparently, the 511~keV data tend to favour such morphologies.
This is illustrated in Fig.~\ref{fig:mlr-bd} where we plot 
the \rmlr\ as function of the bulge-to-disk flux ratio of the tracer map, 
defined as the flux contained within a circular region of $6\deg$ in radius 
around the galactic centre, divided by the flux within galactic latitudes 
$b = \pm 20\deg$ outside the circular bulge region (note that the 
precise B/D value depends of course on the exact definitions chosen 
for the two regions, but our purpose here is to illustrate a trend).
Clearly, there is a strong correlation between the B/D flux ratio and the 
\rmlr, in the sense that the larger the B/D flux ratio, the larger the \rmlr.
In particular, the DIRBE near-infrared maps and the HEAO-1 map show 
the largest B/D flux ratios of all tracer maps ($\sim0.2$).
Thus, finding a population of objects which show a large B/D ratio 
could provide the key for finding the galactic source of positrons.

\subsection{Point-source search}
\label{sec:point}

The modest angular resolution of SPI of about $3\deg$
(Vedrenne et al.~\cite{vedrenne03})
makes it difficult to distinguish between point source, point-like, 
and small-scale diffuse emission.
So in principle we cannot exclude the possibility that the 511~keV gamma-ray 
line emission that is seen towards the galactic bulge region is made of a 
limited number of point sources that blend to simulate diffuse emission.
In due course the question of any point source contribution to the flux 
will be best addressed using data from the imager IBIS on INTEGRAL, which is 
relatively insensitive to diffuse emission, in conjunction with that from SPI.
Such work is underway and will be reported on separately.  
We here limit ourselves to the constraints which can be placed from SPI data 
alone on such a contribution.

We have therefore used the SPIROS algorithm 
(Skinner \& Connell \cite{skinner03}) 
to search our dataset for the positions and fluxes of point sources that 
are compatible with our data.
SPIROS searches for the most probable position for a point source and fits 
a source at that position before repeating iteratively the search using the 
residuals after sources already found are taken into account. 
At each iteration the positions and fluxes of all sources that have been 
found are optimised by maximising a goodness of fit parameter 
(the $\chi^2$ statistic was used here).

If the 511~keV emission is intrinsically diffuse then application of this 
algorithm will lead to sources being placed at selected positions (regions 
of high flux and local noise peaks) until a distribution of emission is 
found that is consistent (given limited statistics) with the data.
Such a description is unlikely, however, to be unique and if most or all 
of the flux is from diffuse emission then the particular source positions 
found will have no astrophysical significance. 
We therefore do not present here the detailed results of this blind 
source search and we restrict ourselves to discussion of the general 
conclusions which can be drawn from the analysis. 

For models with 7 point sources based on the SPIROS solution after 
iteration 7 we obtain $\rmlr = 462.5$, slightly inferior to our best 
fitting diffuse models.
Subsequent iterations suggest point sources at the edge of the 
exposed regions which are obviously spurious.
Only after iteration 13 is another point source found in the galactic 
bulge region.
Fitting this source together with the 7 sources found earlier
leads to $\rmlr = 471.9$, comparable to the best fitting diffuse 
models.
The total flux attributed to the 8 point sources is $1.1\times 10^{-3}$ \funit, 
comparable with the values obtained for the bulge component using our best 
fitting diffuse models (c.f.~Table \ref{tab:modelfit}). 
We therefore simply conclude that at least 8 point sources would be needed to 
satisfactorily describe the SPI data.\footnote{
Given the angular resolution of SPI, compact sources with an extent 
$\sim1\deg-2\deg$ would be considered as point sources in this context.}

\begin{table*}[ht!]
  \footnotesize
  \caption{\label{tab:limits}
  511 keV narrow line $3\sigma$ upper flux limits for selected
  potential positron sources.
  }
  \begin{flushleft}
	\begin{tabular}{lrrccl}
	\hline
	\noalign{\smallskip}
	Source name & \multicolumn{1}{c}{l} & \multicolumn{1}{c}{b} & FWHM &
	$3\sigma$ flux limit & Type of object \\
	&      (deg) &  (deg) & (deg)  & ($10^{-4}$ \funit) &  \\
	\noalign{\smallskip}
	\hline
	\noalign{\smallskip}
	Sgr A$^{\ast}$  &   $0.00\deg$ &  $+0.00\deg$ & - & $<1.0$ & BHC \\
	GRS 1758-258    &   $4.51\deg$ &  $-1.36\deg$ & - & $<0.7$ & $\mu$QSO \\
	Cyg X-1         &  $71.33\deg$ &  $+3.07\deg$ & - & $<1.0$ & BHC \\
	LS I+61\deg303  & $135.68\deg$ &  $+1.09\deg$ & - & $<3.3$ & $\mu$QSO \\
	1E 1740.7-2942  & $359.15\deg$ &  $-0.12\deg$ & - & $<0.9$ & $\mu$QSO \\
	\noalign{\smallskip}
	\hline
	\noalign{\smallskip}
	Vela X-1        & $263.06\deg$ &  $+3.93\deg$ & - & $<1.1$ & HMXB \\
	\noalign{\smallskip}
	\hline
	\noalign{\smallskip}
	GX 5-1          &   $5.08\deg$ &  $-1.02\deg$ & - & $<0.7$ & LMXB \\
	GRS 1915+105    &  $45.37\deg$ &  $-0.22\deg$ & - & $<1.0$ & LMXB \\
	A 0620-00       & $209.96\deg$ &  $-6.54\deg$ & - & $<3.8$ & LMXB \\
	Nova Muscae     & $295.30\deg$ &  $-7.07\deg$ & - & $<2.0$ & LMXB \\
	Cir X-1         & $322.12\deg$ &  $+0.04\deg$ & - & $<1.1$ & LMXB \\
	Cen X-4         & $332.24\deg$ & $+23.89\deg$ & - & $<1.7$ & LMXB \\
	GX 349+2        & $349.10\deg$ &  $+2.75\deg$ & - & $<0.8$ & LMXB \\
	Sco X-1         & $359.09\deg$ & $+23.78\deg$ & - & $<1.5$ & LMXB \\
	\noalign{\smallskip}
	\hline
	\noalign{\smallskip}
	Crab            & $184.56\deg$ & $-5.78\deg$ & - & $<1.3$ & Pulsar \\
	Geminga         & $195.13\deg$ & $+4.27\deg$ & - & $<2.2$ & Pulsar \\
	PSR J0737-3039  & $245.24\deg$ & $-4.50\deg$ & - & $<3.0$ & Pulsar \\
	\noalign{\smallskip}
	\hline
	\noalign{\smallskip}
	Kepler          &   $4.52\deg$ &  $+6.82\deg$ & - & $<0.7$ & SNR \\
	Cas A           & $111.73\deg$ &  $-2.13\deg$ & - & $<1.9$ & SNR \\
	Tycho           & $120.07\deg$ &  $+1.44\deg$ & - & $<2.0$ & SNR \\
	SN 1987A        & $279.70\deg$ & $-31.94\deg$ & - & $<1.0$ & SNR \\
	SN 1006         & $327.58\deg$ & $+14.59\deg$ & - & $<1.3$ & SNR \\
	Lupus Loop      & $329.80\deg$ & $+16.00\deg$ & - & $<1.3$ & SNR \\
	\noalign{\smallskip}
	\hline
	\noalign{\smallskip}
	Cygnus          &    $79\deg$ &    $0\deg$ & $5\deg$ & $<1.8$ & Star forming region (\al\ source) \\
	Vela            &   $265\deg$ &    $0\deg$ & $5\deg$ & $<2.6$ & Star forming region (\al\ source) \\
	Carina          & $286.5\deg$ & $+0.5\deg$ & -       & $<2.2$ & Star forming region (\al\ source) \\
	\noalign{\smallskip}
	\hline
	\noalign{\smallskip}
	M 22            &   $9.89\deg$ &  $-7.55\deg$ & - & $<0.8$ & Globular Cluster \\
	Palomar 13      &  $87.10\deg$ & $-42.70\deg$ & - & $<7.0$ & Globular Cluster \\
	$\omega$ Cen    & $309.10\deg$ & $+14.97\deg$ & - & $<1.7$ & Globular Cluster \\
	NGC 6397        & $338.16\deg$ & $-11.96\deg$ & - & $<1.2$ & Globular Cluster \\
	M 4             & $350.97\deg$ & $+15.97\deg$ & - & $<1.1$ & Globular Cluster \\
	\noalign{\smallskip}
	\hline
	\noalign{\smallskip}
	Sgr dwarf       &   $5.61\deg$ & $-14.10\deg$ &    $8\deg$ & $<1.7$ & Dwarf Galaxy \\
	M31             & $121.17\deg$ & $-21.57\deg$ &    $3\deg$ & $<11.8$ & Galaxy \\
	LMC             & $280.47\deg$ & $-32.89\deg$ & $10.8\deg$ & $<4.7$ & Irregular Galaxy \\
	\noalign{\smallskip}
	\hline
	\noalign{\smallskip}
	QSO B2251-179   &  $46.20\deg$ & $-61.33\deg$ & - & $<2.9$ & QSO \\
	3C 273          & $289.95\deg$ & $+64.36\deg$ & - & $<1.2$ & QSO \\
	3C 279          & $305.10\deg$ & $+57.06\deg$ & - & $<1.3$ & Blazar \\
	Cen A           & $309.52\deg$ & $+19.42\deg$ & - & $<1.7$ & Radio Galaxy \\
	\noalign{\smallskip}
	\hline
	\noalign{\smallskip}
	Coma cluster    &  $58.08\deg$ & $+87.96\deg$ &  $4\deg$ & $<2.0$ & Galaxy Cluster \\
	Perseus cluster & $150.58\deg$ & $-13.26\deg$ &  $4\deg$ & $<4.5$ & Galaxy Cluster \\
	Virgo cluster   & $281.63\deg$ & $+75.18\deg$ & $12\deg$ & $<2.5$ & Galaxy Cluster \\
	\noalign{\smallskip}
	\hline
	\end{tabular}
  \end{flushleft}
\end{table*}

In addition to the blind search for point sources, we also looked for 
evidence of 511 keV gamma-ray line emission from a list of potential 
candidate objects.
Our list comprises compact objects, pulsars, supernova remnants, star 
forming regions, globular clusters, nearby (active) galaxies, and galaxy 
clusters.
Depending on the expected source extent, we searched for either point 
source emission or extended emission, modelled by a 
2d angular Gaussian surface brightness distribution
for which we specify the centroid and the FWHM extension.

The results of the analysis are summarised in Table \ref{tab:limits}.
None of the sources we searched for showed a significant 511~keV 
flux, hence we only quote ($3\sigma$) upper limits in 
Table \ref{tab:limits}.
Since the emission of the Galaxy may interfere with the emission from 
the specific sources (due to the large field of view of the SPI 
instrument), we also included models for the diffuse galactic 511~keV 
emission for the source search.
We have used combinations Shells+D0, Shells+D1, H'+D0 and H'+D1 to cover 
the range of plausible best fitting diffuse models 
(c.f.~Fig.~\ref{fig:modelmaps}) and quote always the most 
conservative flux limit.
The (less sensitive) \orbit\ background model has been used to ensure 
that systematic uncertainties are negligible.

\section{Discussion}
\label{sec:discussion}

\subsection{Comparison with earlier measurements}

Observations of the 511 keV line emission have been made by a large 
number of balloon and satellite borne telescopes, yet only a few of 
them provided constraining information on the emission morphology.
The OSSE instrument that flew during 1991-2000 on-board CGRO
accumulated so far the largest database for studying the 511 keV line 
intensity distribution.
The observations of the Gamma-Ray Spectrometer on-board the 
Solar Maximum Mission (SMM) (1980-1988) and of the 
Transient Gamma-Ray Spectrometer (TGRS) on-board the WIND mission 
(1995-1997)
have also been used to estimate the overall 511 keV line flux and maximum 
emission size
(Purcell et al.~\cite{purcell94}; 
Kinzer et al.~\cite{kinzer96};
Tueller et al.~\cite{tueller96};
Cheng et al.~\cite{cheng97}; 
Purcell et al.~\cite{purcell97};
Harris et al.~\cite{harris98}; 
Milne et al.~\cite{milne00};
Kinzer et al.~\cite{kinzer01}).

The picture that emerged prior to the INTEGRAL launch was 
the following.
From the general trend that instruments with larger fields-of-view show 
larger fluxes it was inferred that the 511 keV emission is extended.
OSSE observations strongly exclude a single point-source located at the 
GC (Purcell et al.~\cite{purcell94}).
The OSSE observations suggest at least two emission components, 
one being a spheroidal bulge and the other being a galactic disk component.
A third component, named the Positive Latitude Enhancement (PLE), situated 
about $9\deg-12\deg$ above the GC has been reported 
(Cheng et al.~\cite{cheng97}; Purcell et al.~\cite{purcell97}), but the 
morphology and intensity of this component was in fact only poorly 
determined by the data
(Milne et al.~\cite{milne00}).

The emission, which showed no significant offset from the GC, was well 
fitted by either a model comprising a narrow ($5-6\deg$ FWHM) 
Gaussian bulge plus $\sim35\deg$ FWHM Gaussian and CO-like disk 
components, or by a centre-truncated R$^{1/4}$ spheroid plus 
exponential disk model 
(Purcell et al.~\cite{purcell97}; Milne et al.~\cite{milne00};
Kinzer et al.~\cite{kinzer01}).
The total 511 keV gamma-ray line flux was estimated to be
$(1-3) \times 10^{-3}$ \funit.
The distribution of flux between the bulge and disk components was 
only weakly constrained by the observations, and depended sensitively 
on the assumed bulge shape.
In particular, estimates for the bulge-to-disk (B/D) flux ratio 
varied from $0.2-3.3$ depending upon whether the bulge component 
features a halo (which leads to a large B/D ratio) or not 
(Milne et al.~\cite{milne00}).

\begin{table}[!t]
  \footnotesize
  \caption{\label{tab:comparison}
  Comparison of SPI results with OSSE measurements.
  The bulge parameters $l_0$, $b_0$, $\Delta l$, and $\Delta b$ of 
  OSSE were taken from Kinzer et al.~(\cite{kinzer01}).
  }
  \begin{flushleft}
  \begin{tabular}{lcc}
  \hline
  \hline
  \noalign{\smallskip}
  Quantity & SPI & OSSE \\ 
  \noalign{\smallskip}
  \hline
  \noalign{\smallskip}
  $l_0$             & $-0.6\deg \pm 0.3\deg$ & $-0.25\deg \pm 0.25\deg$ \\
  $b_0$             & $+0.1\deg \pm 0.3\deg$ &  $-0.3\deg \pm 0.2\deg$ \\
  $\Delta l$ (FWHM) & $8.1\deg \pm 0.9\deg$  & $6.3\deg \pm 1.5\deg$ \\
  $\Delta b$ (FWHM) & $7.2\deg \pm 0.9\deg$  & $4.9\deg \pm 0.7\deg$ \\
  Flux ($10^{-3}$ \funit) & $1.5-2.9$ & $1-3$ \\
  B/D flux ratio    & $1-3$ & $0.2-3.3$ \\
  \noalign{\smallskip}
  \hline
  \end{tabular}
  \end{flushleft}
\end{table}

Our analysis basically confirms the pre-INTEGRAL observations 
(c.f.~Table \ref{tab:comparison}).
One difference is that the bulge appears slightly larger in our analysis 
when compared to the OSSE result.
We note that OSSE performed differential measurements using its
$4\deg \times 11\deg$ collimator which may bias the results towards 
small values (Kinzer et al.~\cite{kinzer01}), but in any case, the 
discrepancy is not very significant and is not surprising in 
view of possible systematic uncertainties.

Another difference with respect to OSSE is that we find no evidence 
for a feature resembling the PLE.
Fitting a model of the PLE 
(2d Gaussian of $5\deg$ FWHM located at $l=-2\deg$ and $b=8\deg$)
on top of the bulge results in a $3\sigma$ upper flux limit of 
$1.5 \times 10^{-4}$ \funit\ for the PLE.
The OSSE team has gradually reduced their estimates of the flux 
and significance attributed to this emission feature from
$5 \times 10^{-4}$ \funit\ (Purcell et al.~\cite{purcell97})
down to an upper limit of
$1 \times 10^{-4}$ \funit\ (Milne et al.~\cite{milne01}).
Recently it has been suggested that data analysis problems linked 
with variable continuum emission may account for the reported PLE
(Milne \cite{milne04}), so perhaps our non-detection of a 
PLE feature is not surprising.

Until now, there have been very few published upper limits on 511 keV 
gamma-ray line emission from point sources which take account of 
diffuse emission.
Examples are the $3\sigma$ limits of
$1.6 \times 10^{-4}$ \funit\ for 1E~1740.7-2942
(Purcell et al.~\cite{purcell94}) and
$\sim1.4 \times 10^{-4}$ \funit\ for inner Galaxy sources
(Milne et al.~\cite{milne01}).
Our upper limits, summarised in Table \ref{tab:limits}, are 
somewhat more stringent.

Finally, we want to mention that the method of analysis used in this 
work assumes that the 511~keV line emission is not time variable. 
From our analysis of SPI data alone we have no indications for time 
variability.
Furthermore, OSSE and TGRS measurements revealed no significant time 
variability (Purcell et al.~\cite{purcell97}; Harris et al.~\cite{harris98}) 
and in addition, our 511~keV line flux measurements are consistent with 
those of OSSE and TGRS.
Thus we believe that our assumption of non-variable 511~keV line emission 
is reasonable.
We have, however, not yet performed a thorough analysis on all 
relevant timescales.

\subsection{General considerations}

The most distinctive morphological feature of the 511~keV emission is 
the large B/D luminosity ratio of $3-9$.
Unless there is a mechanism that strongly suppresses positron 
annihilation in the galactic disk, or that somehow transports 
positrons from the disk into the galactic bulge or halo where they 
annihilate, the positron source population we are seeking for should also 
exhibit such a high B/D ratio.

The B/D luminosity ratio of $3-9$ is considerably larger than the B/D 
mass ratio of $0.3-1.0$ of our Galaxy
(e.g.~Caldwell \& Ostriker \cite{caldwell81};
Freudenreich \cite{freudenreich98}; 
Bissantz \& Gerhard \cite{bissantz02};
Robin et al.~\cite{robin03}).
The uncertainty in the galactic B/D mass ratio is partly due to differences 
in the modelling of the disk component, where disk profiles exhibiting a 
central hole or depletion lead to B/D ratios at the high end, while 
double exponential profiles without hole favour B/D ratios at the low end.
Since we employed in our analysis disk models with central holes from 
Robin et al.~(\cite{robin03}), we should for consistency compare our 
511~keV B/D luminosity ratio to their (large) B/D ratio of 
$\sim1$.\footnote{
We note that, independently, large B/D ratios are also favoured by microlensing 
surveys towards the galactic bulge region (Binney \& Evans \cite{binney01}).
}
But even with such large B/D mass ratios, the source population we 
are seeking for should still be at least 3 times more abundant in the bulge than 
in the disk of the Galaxy.
We therefore conclude that the primary positron source of the Galaxy is 
clearly associated with the galactic bulge.
It therefore should belong to the old stellar population.

Furthermore, the fact that the 511~keV emission matches well the morphology 
of the stellar bulge suggests that positron diffusion probably plays only a 
minor role.
Were positron diffusion to be important we would expect to find substantial 
511~keV emission in gas-rich regions adjacent to the rather gas-poor
galactic bulge, such as the molecular ring structure at galactocentric 
distances of $\sim4$ kpc.
However, we do not find any evidence for 511~keV emission correlated with 
this structure.
We therefore conclude that positron diffusion is negligible at 
galactic scales (i.e.~kpc scales).

\subsection{Constraints on the disk source}
\label{sec:disk}

One certain source of positrons in the disk of the Galaxy is the 
radioisotope \al.
It decays with a lifetime of $\tau \sim 10^6$ yr with emission of a 
1809 keV gamma-ray photon; $\sim85\%$ of the decays are also accompanied 
by the emission of a positron.
The galactic distribution of \al\ is well known thanks to observations 
of the COMPTEL telescope aboard CGRO, and follows that of the young 
stellar population.
Thus, under the assumption that the positrons annihilate close to their 
production site, 511~keV line emission along the galactic plane is 
expected, showing the morphological characteristics of a young stellar 
population.

The expected 511~keV line flux $F_{511}$ due to \al\ decay is related 
to the 1809~keV line flux $F_{1809}$ through
$F_{511} = 0.85 \times (2 - 1.5 \fpositron) \times F_{1809}$.
Using the COMPTEL measurement of the 1809 keV flux along the galactic 
plane,
$F_{1809} = 9 \times 10^{-4}$ \funit\ 
(c.f.~Table 4.3 in Kn\"odlseder \cite{knoedl97}),
and assuming $\fpositron = 0.93$
leads to an expected 511~keV line flux of 
$F_{511} = 5 \times 10^{-4}$ \funit.

Fitting our model of the young stellar population (model D0), together 
with bulge models to the data suggests a disk flux in the range
$(4-6) \times 10^{-4}$ \funit.
To explore the sensitivity of the disk flux on the assumed disk model 
we also paired the shell model with tracers of 1809~keV line emission, 
such as the DMR free-free and the DIRBE 240 \um\ emission maps
(Kn\"odlseder et al.~\cite{knoedl99b}).
This resulted in slightly larger (and more significant) disk fluxes of
$(8.3 \pm 2.3) \times 10^{-4}$ \funit.
Comparing these values to what we expect from \al\ suggests that 
$60-100\%$ of the galactic plane emission may be attributed 
to $\beta^+$-decay of \al.

If this contribution is subtracted a 511~keV disk flux of at 
most $\sim 3 \times 10^{-4}$ \funit\ remains to be explained by other 
positron sources.
A possible candidate is the radioisotope \ti\ whose 
daughter isotope $^{44}$Sc decays via $\beta^+$-decay into stable 
$^{44}$Ca ($\tau \sim 87$ yr).
In contrast to \al\ there is no firm measurement of the present day 
galactic \ti\ mass (nor of its spatial distribution), but simple chemical 
evolution arguments lead to the expectation that about 
$4 \times 10^{-6}$ \Msol\ of \ti\ are produced per year 
(Leising \& Share \cite{leising94}).
Under the assumption that all positrons escape the production site 
this yield translates into an annihilation rate of 
$3 \times 10^{42}$ \peunit.
Assuming further that \ti\ is distributed following model D0, a 
511~keV disk flux of $\sim 8 \times 10^{-4}$ \funit\ is expected.
In view of the approximate estimation and in particular in view of the 
uncertainty about the spatial distribution this value seems in 
reasonable agreement with the remaining 511~keV flux of 
$\sim 3 \times 10^{-4}$ \funit.

It is intriguing that the galactic disk flux could be entirely explained 
by the radioactive decay of \al\ and \ti.
This would suggest that once the \al\ and \ti\ contributions have been 
subtracted only the bright bulge component of 511~keV emission remains, 
which would then demand a specific source population that is only confined 
to the inner Galaxy.
However, we cannot immediately draw this conclusion.
Fitting an old stellar population disk model, with larger scale height 
and smaller radial scale than the young one, increases the estimate of 
the disk flux by about a factor of two, leaving room for a weak disk 
component not associated with \al\ or \ti.
This would suggest lower limits on the B/D flux (luminosity) ratio of 
$\ga2$ ($\ga6$) for the source population that gives rise to the
galactic bulge emission.

\subsection{Constraints on the bulge source}

\subsubsection{Massive stars, core collapse supernovae, pulsars}
\label{sec:massive}

The bulge dominance of the 511~keV emission immediately excludes 
scenarios in which the bulk of galactic positron production is related to 
massive stars.
Such scenarios include the production of $\beta^+$-decay radioisotopes 
produced by Wolf-Rayet stars and all types of core collapse supernovae 
(including hypernovae) and the pair production in the strong magnetic fields 
of pulsars.
Massive stars may well explain the faint disk component of 511~keV emission 
via the radioactive decay of \al\ and \ti\ (c.f.~section \ref{sec:disk}).
They cannot, however, explain the majority of the emission, which 
would in that case resemble the 1809 keV line emission.

\subsubsection{Hypernovae}

Cass\'e et al.~(\cite{casse04}) proposed that a recent 
hypernova at the galactic centre could be responsible for the observed 
positron emission, but there is no observational evidence that such an 
explosion indeed took place.
Hypernovae are believed to be related to Wolf-Rayet stars, which are 
distributed in the galactic disk following the spiral arm pattern, 
hence it would be much more likely to find a recent hypernova at an 
arbitrary position along the galactic plane (or at the tangent points 
of the spiral arms) rather than at the position of the galactic centre.
And even in the rare event of a hypernova exploding right at the galactic 
centre, it would be difficult to explain why the resulting 511~keV 
annihilation radiation (arising from the $\beta^+$-decay of freshly 
synthesised $^{56}$Co) should reflect the stellar morphology of the 
galactic bulge.
We therefore conclude that it is unlikely that galactic centre 
hypernovae are the source of the bulge positrons.

\subsubsection{Cosmic-ray interactions}

Interactions of cosmic-ray particles with the ambient interstellar 
medium may produce positrons, primarily via the
${\rm N} + {\rm p} \to \pi^+ \to {\rm e}^+$
reaction channel (N stands for nucleus).
The effect of the cosmic-ray interaction is best seen in the GeV 
gamma-ray domain, and has been comprehensively mapped by the EGRET 
satellite aboard CGRO.
The EGRET all-sky map shows dominant emission from the galactic 
plane, which follows a linear combination of various gas and dust 
tracers in the galaxy.
Thus cosmic-ray interactions should also lead to disk dominated
511~keV emission, which is at odds with our observations.

\subsubsection{X-ray binaries}
\label{sec:lmxb}

Positron production in X-ray binaries may occur either as a result of
$\gamma\gamma$ pair production in the luminous compact region around 
the compact object or due to nuclear interactions that may form excited 
nuclei that subsequently decay through the emission of positrons.
Galactic black holes and microquasars, where the positrons are ejected in 
a relativistic jet with Lorentz factors of a few 
(Dermer \& Murphy \cite{dermer01}), are the two leading candidates.

X-ray binaries are separated into two classes, depending on whether 
the donor is a high-mass star (HMXB) or a low-mass star (LMXB).
The two classes show clearly different spatial distributions 
(Grimm et al.~\cite{grimm02}).
HMXB are associated with the young stellar population and are 
primarily found in the galactic disk.
Consequently they can immediately be excluded as the source of the bulge 
positrons.
LMXB, in contrast, are strongly concentrated towards the galactic 
bulge, and are more promising source candidates.
Among the 150 LMXBs listed in the catalogue of Liu et al.~(\cite{liu01}), 
more than $50\%$ are observed towards the galactic bulge.
Correcting for completeness, Grimm et al.~(\cite{grimm02}) find a 
B/D ratio of $\sim0.9$ and a vertical scale height of 410~pc for the 
LMXB distribution.

Formally, the LMXB B/D ratio is considerably below the value required 
by our 511~keV data, yet the large vertical scale height of LMXB 
could lead to a scenario where a substantial fraction of positrons from 
disk LMXB may escape into the galactic halo.
This scenario works as follows.
Since the scale height of LMXB (410~pc) considerably exceeds the scale 
height of the dense interstellar gas layer of our Galaxy ($\sim100$ pc), 
positrons from disk LMXB are ejected into rather low-density regions, 
typically a factor of $10-100$ less dense than regions found near the 
galactic plane (Ferri\`ere \cite{ferriere98}).
Before positrons can annihilate they have to slow down considerably, 
mostly through Coulomb interactions, with a characteristic timescale 
of $\tau_{\rm SD} \sim 10^{5} n^{-1}$ yr, $n$ being the ISM density 
in units of cm$^{-3}$ (Forman et al.~\cite{forman86}).
Consequently, positrons live $10-100$ times longer at large scale heights 
than near the galactic plane, allowing for substantial diffusion before 
annihilation takes place.
The typical diffusion length depends much on the magnetic field 
configuration and the amount of ISM turbulence at large scale heights,
but qualitatively it seems plausible that a considerable fraction 
of the disk positrons may annihilate in the galactic halo.

The resulting broad diffuse component of 511~keV emission would be 
difficult to detect with SPI.
In particular, the present dataset, for which good exposure is 
restricted to a band $b \la \pm15\deg$ along the galactic plane
(c.f.~Fig.~\ref{fig:exposure}), makes it virtually impossible to measure 
a disk component with a broad-latitude distribution of 511~keV emission.
Therefore it would be sufficient that $\sim2/3$ of the positrons produced 
by disk LMXB escape into the galactic halo to reconcile the LMXB 
distribution with a B/D ratio of 3.
If we require a more extreme B/D ratio of 6, as expected after 
subtraction of the \al\ component from the disk
(c.f.~section \ref{sec:disk}), a positron escape fraction of $\sim80\%$ 
would be needed for disk LMXB.
It remains to be seen whether such large escape fractions are feasible.

An alternative way to test the LMXB scenario is to search for 511~keV line 
emission from individual bright and/or nearby objects.
So far no emission is seen towards the interesting candidates
Sco~X-1 (the brightest LMXB) and Cen~X-4 and A0620-00 (probably the most 
nearby LMXB at $\sim1.2$ kpc), but we plan for deep observations of these 
objects in the near future to search for their annihilation signatures.
The detection of positron annihilation signatures from nearby objects could 
however be hampered by (even modest) positron diffusion away from the sources, 
which would lead to extended 511~keV emission halos around the objects.
So even for the modest angular resolution of SPI of $\sim3\deg$ nearby 
individual LMXB could appear as extended sources, and their low
surface brightness could make their detection more difficult.

\subsubsection{Classical novae}

Among all proposed positron candidate sources, classical novae,
i.e.~thermonuclear runaways on white dwarfs in accreting binary systems, 
are the sources for which the largest B/D ratios of $\sim3-4$ have been 
suggested
(Della Valle \& Duerbeck \cite{dellavalle93};
Della Valle \& Livio \cite{dellavalle94a}).
Interstellar extinction, in particular towards the galactic bulge region, 
makes it virtually impossible to derive their spatial distribution in 
the Milky Way directly, but novae are readily observed in nearby external 
galaxies which may serve as templates (e.g.~Shafter et al.~\cite{shafter00}).
Due to its proximity and due to its similarity to the Milky Way, M31 
is the primary source of information, and modern investigations 
indicate that novae reside primarily in the bulge region of M31
(Ciardullo et al.~\cite{ciardullo87};
Capaccioli et al.~\cite{capaccioli89}).
Although it had been suggested that selection effects may have ``faked'' 
such a finding (Hatano et al.~\cite{hatano97})
the recent study of Shafter \& Irby (\cite{shafter01}) demonstrates that 
such biases, if they exist, must be small.

Novae produce positrons via the $\beta^+$-decay of radioactive 
isotopes synthesised during the thermonuclear runaway, mainly of
$^{13}$N, $^{18}$F, and $^{22}$Na
(lifetimes $\tau=14$~min, $2.6$~hr, and $3.75$~yr, respectively).
$^{22}$Na yields of $6 \times 10^{-9}$ \Msol, as suggested by 
theoretical nucleosynthesis calculations for ONe novae
(Hernanz et al.~\cite{hernanz02}), would require nova rates of 
$\sim1600$~yr$^{-1}$ to maintain positron production and 
annihilation in an equilibrium state, a rate which is considerably above 
estimates of $35 \pm 11$~yr$^{-1}$ for all types of novae in the entire 
Galaxy (Shafter \cite{shafter97}).
$^{13}$N yields of $2 \times 10^{-7}$ \Msol\ for low-mass CO novae are 
more promising (Hernanz et al.~\cite{hernanz02}) since they would 
require nova rates of only $26$~yr$^{-1}$ if
all positrons could indeed escape from the nova envelope into the ISM.
However, with a $^{13}$N lifetime of $14$~min it seems unlikely that this 
would be possible.

It is probable that large fractions of the $^{13}$N positrons annihilate 
within the dense nova envelope, leading to prompt annihilation that could 
give rise to transient annihilation signatures 
(Leising \& Clayton \cite{leising87}; G\'omez-Gomar et al.~\cite{gomez98}).
This signature has been sought using various gamma-ray telescopes, 
but has so far eluded detection (see Hernanz \& Jos\'e \cite{hernanz04} and 
references therein).
Detection of the transient signature may help to shed light on 
the positron escape fraction, and could show whether 
novae contribute to the galactic bulge positron budget or not.

\subsubsection{Thermonuclear supernovae}

In view of their potential to produce large numbers of positrons,
thermonuclear Type Ia supernovae (SN~Ia) are often considered as the 
most plausible source of positrons in the Milky Way 
(Dermer \& Murphy \cite{dermer01}).
SN~Ia produce positrons via the $\beta^+$-decay of radioactive $^{56}$Co 
($\tau=111$ days).
Expected $^{56}$Co yields of $\sim0.6$ \Msol\ provide 
$\sim2.5 \times 10^{54}$ positrons per event, although, as with novae, 
prompt annihilation in the supernova envelope probably prevents large 
fractions of the positrons from escaping into the ISM.
From the analysis of late light curves of SN~Ia
Milne et al.~(\cite{milne99}) derive a mean escaped positron yield of
$\sim8 \times 10^{52}$ positrons per SN~Ia,
corresponding to a positron escape fraction of $f\sim0.03$.
A recent study of SN~2000cx even suggests $f\sim0$, but SN~2000cx was 
an unusual event that may not represent the average SN in the bulge of 
our Galaxy (Sollerman et al.~\cite{sollerman04}).

Assuming therefore $f=0.03$ a bulge SN~Ia rate of $0.6$ per century 
is required to maintain the observed 511~keV luminosity in a steady state.
Unfortunately the rate and distribution of SN~Ia in our Galaxy are only 
poorly known.
The galactic SN~Ia rate is generally inferred from rates observed in 
external galaxies which are then scaled to the mass and the type of the 
Milky Way.
In that way rates of
$0.3-1.1$ SN~Ia per century are derived
(Tammann et al.~\cite{tammann94};
Cappellaro et al.~\cite{cappellaro97};
Mannucci et al.~\cite{mannucci05}),
sufficient to maintain the galactic 511~keV luminosity.
In contrast, when we follow the suggestion of 
Prantzos (\cite{prantzos04}) and derive the bulge SN~Ia rate by 
scaling the SN~Ia rate observed in early-type galaxies to the mass of 
the galactic bulge, a bulge SN~Ia rate of
$0.08$ SN~Ia per century is obtained.
This value is much too low to explain the observed bulge 511~keV 
luminosity.
It is difficult to judge if the galactic bulge can indeed be 
considered as a downsized version of an elliptical galaxy, in particular in 
view of the differences in the evolution of the galactic 
bulge and an elliptical galaxy.
Furthermore, it is suggested that different SN~Ia explosion mechanisms 
exist in different types of galaxy
(e.g.~Della Valle \& Livio \cite{dellavalle94b};
Howell \cite{howell01};
Mannucci et al.~\cite{mannucci05})
making the proposed extrapolation even more uncertain.

Observations of external galaxies indicate that SN~Ia 
distributions are strongly peaked towards galactic centres
(Bartunov et al.~\cite{bartunov92}), yet reliable determinations of 
B/D ratios are difficult in view of observational biases and selection 
effects
(Wang et al.~\cite{wang97}; 
Hatano et al.~\cite{hatano98};
Howell et al.~\cite{howell00}).
If the SN~Ia distributions follows that of novae
(both populations are believed to arise from accreting white dwarfs;
see van den Bergh \cite{vandenbergh88}) one can expect B/D ratios of
$\sim3-4$.
Even higher B/D ratios can be achieved if part of the positrons  
produced by disk SN~Ia, which have a vertical scale height of $\sim330$ pc
(Chen et al.~\cite{chen01}), escape into the halo
(c.f.~section \ref{sec:lmxb}).
Thus SN~Ia could indeed present the required characteristics that are 
needed to explain the positron distribution and annihilation rate in the Galaxy.

Alternatively, instead of explaining the bulge emission globally we may 
search for 511~keV emission from nearby Type Ia supernovae remnants, 
such as SN~1006, the Lupus Loop, or the Tycho SNR.
Assuming a mean escaped positron yield of $\sim8 \times 10^{52}$ positrons 
per SN~Ia (Milne et al.~\cite{milne99}) and a
positronium fraction of $\fpositron=0.93$, the mean
expected 511~keV line flux from an individual SN~Ia is estimated to
\begin{equation}
 F_{511} = 1.3 \times 10^{-4} \left( \frac{1~{\rm kpc}}{D} \right)^2
			      \left( \frac{10^5~{\rm yr}}{\tau} \right) 
			      \funit\
\end{equation}
where $D$ is the distance to the supernova remnant in units of kpc, and 
$\tau$ is the mean lifetime of the positrons in years, 
between $10^3-10^7$ yr, depending on the density, temperature, 
and ionisation state of the annihilating medium 
(Guessoum et al.~\cite{guessoum91}).
Taking distances
of $D\sim2$ kpc for SN~1006 (Laming et al.~\cite{laming96}),
of $D\sim1$ kpc for the Lupus Loop (Leahy et al.~\cite{leahy91}), 
and of $D\sim2.3$ kpc for Tycho (Hughes \cite{hughes00})
and assuming a positron lifetime of $10^5$ yr 
results in predicted 511~keV line fluxes of
$0.3 \times 10^{-4}$ \funit,
$1.3 \times 10^{-4}$ \funit, and
$0.2 \times 10^{-4}$ \funit, respectively.
At least for the Lupus Loop, the predicted flux is close to our upper 
511~keV flux limit, indicating that dedicated deep observations of 
nearby supernova remnants can help to answer the question about the 
galactic positron source.
Such dedicated observations with INTEGRAL are already scheduled.

\subsubsection{Light dark matter annihilation}

Light dark matter (1-100 \MeV) annihilation, as suggested recently by 
Boehm et al.~(\cite{boehm04}),
is probably the most exotic but also the most exciting candidate source 
of galactic positrons.
Unfortunately, the spatial distribution of dark matter in general, and 
light dark matter in particular, is only poorly constrained by 
observational data, at least for the inner Galaxy.
The debate of whether the dark matter profile shows a cusp towards the 
galactic centre is still not settled, but it seems clear now that, 
dynamically, dark matter plays only a minor role in the inner 3~kpc 
of the Galaxy. 
In this region the stellar mass dominates
(Binney \& Evans \cite{binney01}; Klypin et al.~\cite{klypin02}).

Due to these uncertainties it is difficult to judge whether the 
observed 511~keV emission could be explained by dark matter 
annihilation.
Maybe more promising is the idea to search for signatures of dark 
matter annihilation in nearby, external galaxies.
Hooper et al.~(\cite{hooper04}) suggested that nearby dwarf spheroidal 
galaxies may provide prominent sources of 511~keV line emission due 
to the high densities of dark matter that are known to be present.
They proposed the nearby Sagittarius dwarf galaxy (Sgr dwarf) as most 
promising candidate and estimate 511~keV line fluxes in the range
of $(1-7) \times 10^{-4}$ \funit, depending on the assumed dark 
matter halo profile.
Cordier et al.~(\cite{cordier04}) searched for emission from this 
Galaxy using SPI and obtained a $(3\sigma)$ upper limit of
$3.8 \times 10^{-4}$ \funit.
Our upper limit of $1.7 \times 10^{-4}$ \funit\ is substantially lower,
and excludes almost all types of halo models for this galaxy, in 
particular those with a central cusp.

Standard cold dark matter cosmology predicts cuspy dark matter 
distributions (Klypin et al.~\cite{klypin02}), so in principle Sgr dwarf 
should have been detected by SPI if dark matter annihilation were a viable 
scenario.
Maybe dark matter halos are less cuspy than theory predicts?
This possibility is indeed indicated by observations of our own Galaxy
(Binney \& Evans \cite{binney01}) and dwarf galaxies
(Blais-Ouellette et al.~\cite{blais99}; Kleyna et al.~\cite{kleyna03}).
But in this case dark matter annihilation should not lead to a compact 
but to a rather extended 511~keV emission feature -- in contradiction to 
what SPI observations of the inner Galaxy suggest.
From the arguments given one may question the dark matter scenario.
However, it is certainly premature to reject them totally because of 
the uncertainties in the dark halo profiles and the annihilation 
conditions.

\section{Conclusions}
\label{sec:conclusions}

Our first mapping of 511~keV gamma-ray line emission over a large 
fraction of the celestial sphere leads us to the following observations:

\begin{enumerate}
\item 511~keV emission is significantly ($\sim50\sigma$) detected towards 
      the galactic bulge region, and, at a very low level ($\sim4\sigma$), 
      from the galactic disk
\item there is no evidence for a point-like source in addition to the 
      diffuse emission, down to a typical flux limit of
      $\sim10^{-4}$ \funit\
\item there is no evidence for the positive latitude enhancement that 
      has been reported from OSSE measurements;
      the $3\sigma$ upper flux limit for this feature is
      $1.5 \times 10^{-4}$ \funit\
\item the bulge emission is spherically symmetric and is centred on the 
      galactic centre with an extension of $\sim8\deg$ (FWHM);
      it is equally well described by models that represent the stellar 
      bulge or the halo populations
\item the bulge annihilation rate is
      $(1.5 \pm 0.1) \times 10^{43}$ \peunit, 
      the disk annihilation rate is
      $(0.3 \pm 0.2) \times 10^{43}$ \peunit
\item the bulge-to-disk luminosity ratio lies in the range $3-9$
\end{enumerate}

The bulge dominated 511~keV line emission morphology suggests an old 
stellar population as the main galactic positron source.
In contrast, the faint disk emission is well explained by the release 
of positrons during the radioactive decay of \al\ that originated from 
massive stars, with a possible contribution from \ti\ synthesised 
during supernova explosions.

The extreme bulge-to-disk ratio that is observed in the 511~keV 
luminosity imposes severe constraints on the principal galactic positron 
source.
Type Ia supernovae, low-mass X-ray binaries or dark matter annihilation may 
possibly satisfy these constraints, but uncertainties in the knowledge about 
the spatial distribution of these objects and the positron escape processes 
prevents us from drawing firm conclusions.
Novae could probably most easily explain the large B/D ratios, yet an 
implausibly large positron escape fraction from $^{13}$N decay would be 
required to accommodate the observed annihilation rate.
SN~Ia could explain the annihilation rate for a modest positron 
escape fraction, but it is questionable if they have the required large B/D 
ratio.
LMXB could reproduce the observed B/D ratio provided that a 
substantial fraction of positrons ejected by disk LMXB escape into the 
halo.
Light dark matter is an exciting option, but it remains to be seen if 
the observed 511~keV emission distribution is compatible with the
profile of the galactic dark matter halo.

Future deep observations of individual nearby candidate sources may 
provide the means to identify the galactic positron source.
As such we will soon observe the X-ray binaries Sco~X-1 and Cen~X-4 
with INTEGRAL, and observations of the nearby supernova remnant SN~1006 
are already scheduled.
We cannot be sure that any of these observations will allow the 
detection of a 511~keV signal, but were such a signal detected we 
would gain important new insights in the primary source of positrons 
in our Galaxy.

\appendix
\section{Summary of galactic density profiles}
\label{sec:gal-models}

\paragraph{Bulge models:}
We model the galactic bulge as a triaxial stellar bar for which the 
parameters are summarised in Table~\ref{tab:bulgemodel}.
The apparent intensity distribution on the sky is computed in the 
galactic frame which we define by a right-handed cartesian coordinate 
system where the Sun is located on the negative y-axis, at 
$-\Rsol=8.5$ kpc.
The transformation from the galactic frame into the bar frame is 
performed by two consecutive rotations:
the first, represented by the matrix $R_{X}(\alpha)$, is a 
counterclockwise rotation by an angle $\alpha$ around the z-axis;
the second, represented by the matrix $R_{Y}(-\beta)$, is a clockwise 
rotation by an angle $\beta$ around the new y-axis, i.e.
\begin{equation}
 \vec{r'} = R_{Y}(-\beta) R_{X}(\alpha) \vec{r}
\end{equation}
The effective bar radius is defined by
\begin{equation}
 R_{\rm s} = \left( \left[ \left( \frac{|x'|}{a_{\rm x}} \right)^{C_{\perp}} +
                       \left( \frac{|y'|}{a_{\rm y}} \right)^{C_{\perp}}
		\right]^{\frac{C_{\parallel}}{C_{\perp}}} +
		\left( \frac{|z'|}{a_{\rm z}} \right)^{C_{\parallel}}
	 \right)^{\frac{1}{C_{\parallel}}}
\end{equation}
where 
$a_{\rm x}$, $a_{\rm y}$, and $a_{\rm z}$ are the scale lengths and
\cface\ and \cedge\ are the face-on and the edge-on shape parameters.
The radial dependencies of the bar density are given by various 
density profiles that we designate by labels:
\begin{eqnarray}
{\rm G:}  & \rho_{\rm G}  = & \rho_{0} \times f_{\rm max}(R_{\rm xy}) \times 
            \exp(-0.5 R_{\rm s}^2) \\
{\rm G3:} & \rho_{\rm G3} = & \rho_{0} \times f_{\rm max}(R_{\rm xy}) \times 
            R_{\rm s}^{-1.8} \exp(-R_{\rm s}^3) \\
{\rm E:}  & \rho_{\rm E}  = & \rho_{0} \times f_{\rm max}(R_{\rm xy}) \times  
            \exp(-R_{\rm s}^n) \\
{\rm E3:} & \rho_{\rm E3} = & \rho_{0} \times f_{\rm max}(R_{\rm xy}) \times  
            K_{0}(R_{\rm s}) \\
{\rm S:}  & \rho_{\rm S}  = & \rho_{0} \times f_{\rm max}(R_{\rm xy}) \times  
            {\rm sech}^2(R_{\rm s}) \\
{\rm P:}  & \rho_{\rm P}  = & \rho_{0} \times f_{\rm max}(R_{\rm xy}) \times  
            [1+(R_{s}/R_{c})]^{-1}
\end{eqnarray}
where
\begin{eqnarray}
f_{\rm max}(R_{\rm xy}) = \left\{
\begin{array}{lll}
& 1.0 & \mbox{for $R_{\rm xy} \le R_{\rm max}$} \\
& \exp \left(-\frac{1}{2} 
  \left(\frac{R_{\rm xy}-R_{\rm max}}{a_{\rm max}} \right)^2 \right)
& \mbox{for $R_{\rm xy} > R_{\rm max}$}
\end{array} 
\right.
\end{eqnarray}
is a cutoff function and 
\begin{equation}
R_{\rm xy} = \sqrt{x^2 + y^2}
\end{equation}
is the distance from the galactic centre in the xy-plane.

\begin{table*}[!ht]
  \footnotesize
  \caption{\label{tab:bulgemodel}
  Parameters of the triaxial galactic bulge models used in this work 
  (see text).
  }
  \begin{flushleft}
  \begin{tabular}{lccccccccccc}
    \hline
    \hline
    \noalign{\smallskip}
    Model & $\alpha$ & $\beta$ & $a_{x}$ & $a_{y}$ & $a_{z}$ &
    \cface\ & \cedge\ & $n$ & $R_{c}$ & $R_{max}$ & $a_{max}$ \\
    \noalign{\smallskip}
    \hline
    \noalign{\smallskip}
    G0 & $0.0$ & $0.0$ & \multicolumn{2}{c}{$0.91$} & $0.51$
      & $2$ & $2$ & & & $\infty$ & \\
    G1 & $73.8$ & $0.6$ & $3.11$ & $0.76$ & $0.48$
      & $2$ & $2$ & & & $2.40$ & $0.71$ \\
    G2 & $70.7$ & $0.8$ & $1.56$ & $0.60$ & $0.45$
      & $2$ & $4$ & & & $2.40$ & $0.71$ \\
    G3 & $59.0$ & $0.7$ & $3.78$ & $1.44$ & $1.19$
      & $2$ & $2$ & & & $2.40$ & $0.71$ \\
    E0 & $0.0$ & $0.0$ & \multicolumn{2}{c}{} &
      & $2$ & $1$ & & & $\infty$ & \\
    E1 & $65.7$ & $0.5$ & $2.23$ & $0.65$ & $0.32$
      & $1$ & $1$ & & & $2.40$ & $0.71$ \\
    E2 & $49.0$ & $0.6$ & $0.75$ & $0.19$ & $0.27$
      & $2$ & $2$ & & & $2.40$ & $0.71$ \\
    E3 & $49.8$ & $0.7$ & $0.69$ & $0.19$ & $0.28$
      & $2$ & $4$ & & & $2.40$ & $0.71$ \\
    S$_{\rm F}$ & $76.2$ & $0.02$ & $1.70$ & $0.64$ & $0.44$
      & $1.57$ & $3.50$ & & & $3.13$ & $0.46$ \\
    E$_{\rm F}$ & $80.5$ & $-0.05$ & $1.89$ & $0.66$ & $0.43$
      & $1.61$ & $3.49$ & $1.44$ & & $3.57$ & $0.56$ \\
    P$_{\rm F}$ & $76.8$ & $0.02$ & $1.81$ & $0.65$ & $0.43$
      & $1.65$ & $3.02$ & $5.04$ & $1.23$ & $2.71$ & $0.88$ \\
    S$_{\rm PR}$ & $79.4$ & $-0.8$ & $1.87$ & $0.56$ & $0.47$
      & $3.49$ & $3.38$ & & & $3.71$ & \\
    \noalign{\smallskip}
    \hline
  \end{tabular}
  \end{flushleft}
\end{table*}

\paragraph{Model D0:}
For the young disk population we use the model proposed by
Robin et al.~(\cite{robin03}) to describe the young (age $<0.15$ Gyr) 
stellar disk population of the Galaxy:
\begin{equation}
 \rho(x,y,z)=\rho_{0} (\exp(-(a / R_0)^2) - \exp(-(a / R_i)^2))
\end{equation}
where
\begin{equation}
 a^2 = x^2 + y^2 + z^2 / \epsilon^2 .
 \label{eq:defa}
\end{equation}
This model presents a truncated exponential disk profile
with
a fixed disk axis ratio of $\epsilon = 0.014$,
a fixed disk scale radius of $R_0 = 5$ kpc, and
a fixed inner disk truncation radius of $R_i = 3$ kpc.
The vertical exponential scale height of the disk is $z_{0} = 70$ pc.

\paragraph{Model D1:}
For the old disk population we use the model proposed by
Robin et al.~(\cite{robin03}) to describe the old (age $7-10$ Gyr)
stellar disk population of the Galaxy:
\begin{eqnarray}
 \rho(x,y,z) & = & \rho_{0} (\exp(-(0.25 + a^2 / R_0^2)^{1/2}) - 
 \nonumber \\
 & & \exp(-(0.25 + a^2 / R_i^2)^{1/2}))
\end{eqnarray}
This model presents a truncated exponential disk profile
with
a fixed disk axis ratio of $\epsilon = 0.0791$,
a fixed disk scale radius of $R_0 = 2.53$ kpc, and
a fixed inner disk truncation radius of $R_i = 1.32$ kpc.
The vertical exponential scale height of the disk is $z_{0} = 200$ pc.

\paragraph{Model H:}
To model the stellar halo we use the general model proposed by
Robin et al.~(\cite{robin03}):
\begin{equation}
 \rho(R,z)=\rho_{0} (a/\Rsol)^{-n}
\end{equation}
where $a$ is defined by Eq.~\ref{eq:defa}
(the flatness of the model is set by the value of the axis ratio
$\epsilon$ in Eq.~\ref{eq:defa}),
and $a \ge a_{c}$, avoiding a singularity at the galactic centre.
$n$ determines the slope of the density profile.

\paragraph{Model S:}
A set of galactocentric nested spherical shells of constant density.
The radii of the shells and their number has been varied in order
to maximise the \mlr, while using the minimum number of shells 
required to satisfactorily describe the data.

\begin{acknowledgements}
The SPI project has been completed under the responsibility and leadership 
of CNES.
We are grateful to ASI, CEA, CNES, DLR, ESA, INTA, NASA and OSTC for 
support.
\end{acknowledgements}


\end{document}